\documentclass[%
 reprint,
 amsmath,amssymb,
 aps,
]{revtex4-2}

\DeclareUnicodeCharacter{2212}{-}
\DeclareUnicodeCharacter{03B7}{ }
\DeclareUnicodeCharacter{03B4}{ }
\usepackage{comment}
\usepackage{graphicx}
\usepackage{chngcntr}
\usepackage{float}
\usepackage{dcolumn}
\usepackage{natbib}
\usepackage{bm}

\begin{document}

\preprint{APS/123-QED}

\title{Hall Sensor Based Measurements of the Magnetic Phase Diagram \\ of Superconducting Vanadium}

\author{Eshan Kemp$^\dagger$}
\author{Daniel P. Newton$^\dagger$}%
 \email{Correspondance: dpnewton@stanford.edu}
\author{Ricky Parada}%
\email{These authors contributed equally to this work}
\affiliation{%
 Department of Physics, Stanford University, Stanford, California 94305
}%

\date{August 10, 2022}

\begin{abstract}
Vanadium is an elemental type-II superconductor which has a pure superconducting phase, a vortex phase, and a non-superconducting phase. We designed and carried out a low-budget experiment in our undergraduate physics lab to measure the magnetic field a close distance from the surface of a pure Vanadium disk sample isothermally with a Hall sensor across a range of different temperatures and external magnetic fields. We then used these measurements to compute the superconducting phase diagram of Vanadium. After completing measurements at eight temperatures ranging from 2.6K to 5K and magnetic fields ranging from -0.2T to 0.2T, we were able to detect the Meissner phase (clearly), and a vortex phase (less clearly). Using the graphs of the hall bar magnetic field versus external magnetic field, we were able to derive a rough estimate for the upper critical field and the lower critical field of Vanadium at the eight temperatures wherein we conducted measurements. The results are in general accord with the phase diagram of Vanadium found in the literature, except that the critical temperature appears to be less than $5.43K$ and $5.13K$ (about $4.8K$). Further, we observed critical fields lower than those found in the literature. We hypothesize this is due mostly to impurities in our Vanadium sample (purchased through Amazon.com), which would reduce the measured critical temperature below that of pure Vanadium, consistent with our results.

\end{abstract}

\maketitle


\section{\label{sec:level1}Introduction}

There have been a wide class of sophisticated and expensive experiments which have very accurately measured the magnetic properties of elemental type-II superconductors \cite{Leplae}, most notably Vanadium \cite{Tilley} \cite{Radebaugh} and Niobium \cite{Beebe}. In particular, coil based magnetic flux measurements \cite{Sekula}, resistance measurements \cite{Wexler}, and heat capacity measurements \cite{Brown} have all been used to precisely map out the phase diagram of elemental superconductors. This paper, by contrasts, describes a low-budget and relatively simple experiment to calculate the phase diagram of Vanadium through a Hall sensor based measurement approach that we designed and carried out in the context of our undergraduate experimental physics class. The magnetic field above a superconducting sample is directly measured, and used to infer the amount of magnetic flux which penetrated the Vanadium sample. These measurements are then used to infer the phase of the Vanadium sample at different temperatures and external magnetic fields. By using a direct method of probing the magnetic flux penetration from measurements of the magnetic field distribution, we were able to obtain reasonable accuracy at very low cost within the constraints of using only equipment readily available in most undergraduate physics labs at many universities. This experiment might motivate future work in undergraduate or graduate physics labs wherein the magnetic field distribution inside and outside a superconducting sample is directly mapped out in both the Meissner phase, the vortex phase, and in a non-superconducting phase without the necessity of purchasing high-cost equipment and thus could be used to introduce undergraduate or beginning graduate students in experimental physics, who typically have not yet been introduced to the beautiful properties of type-II superconductors. 

More concretely, for our experiment we placed a bulk Vanadium disk within a set of eight temperatures between 2.6K and 5K, and a set of over one hundred magnetic field values ranging from -2T to 2T. At each temperature, we isothermally increased the external magnetic field while measuring the resulting magnetic field at a point near the central axis of the Vanadium disk (where the Hall sensor is placed). These measurements were used to derive isothermal magnetization curves, from which we can deduce both the upper and lower critical magnetic field strengths at the given temperature. Performing this measurement at a variety of different temperatures then allowed us to map out the upper and lower critical field strengths as a function of temperature, and hence the phase diagram of the Vanadium sample. Our measurements indicate that the critical temperature of our Vanadium disk sample is $4.8K$, a bit less than the $5.43K$ \cite{Sekula} and $5.13K$ \cite{Wexler} previously measured critical temperatures of pure Vanadium. Furthermore, we observe that the critical fields we measure for a given temperature is less than those found in the literature. We hypothesize that these discrepancies were due partly to the relative simplicity of our design and low-cost of our measurement equipment, but mostly due to impurities in our Vanadium sample which was purchased through Amazon.com. Nonetheless, the measurements were accurate enough to map out the phase diagram of Vanadium in a simple, direct, and inexpensive way.

\section{Theory}

Phenomenologically, a type-II superconductor is a superconductor which experiences three types of phases: a pure superconducting phase, a normal phase, and an intermediate phase which contains a combination of normal and superconducting properties \cite{Tilley}. The intermediate phase is commonly denoted as either the vortex phase or the Abrikosov phase. In the pure superconducting phase, magnetic field lines are completely expelled by the superconducting sample, so that there is no magnetic flux inside the superconductor. In the vortex phase, the superconducting sample contains small quantas of superconducting current loops, called vortices \cite{Ohuchi}. These vortices admit a quantized amount of magnetic flux, meaning that in the vortex phase the superconducting sample admits a partial amount of magnetic flux. In the normal phase, the superconducting sample admits magnetic field lines inside it, and has a magnetic flux determined by the magnetic permeability of the superconducting substance in its normal phase.
    
The specific phase diagram of a type-II superconductor as shown in figure 1 can be expressed as follows. Consider a superconducting sample subject to different temperatures and external magnetic fields. Then, at a given temperature $T$, the upper critical field $B_{c2}(T)$ separates the normal phase (which occurs at external magnetic field strengths greater than $B_{c2}(T)$), from the vortex phase \cite{Iryna, Ma}. The lower critical field $B_{c1}(T)$ separates the vortex phase from the normal phase (which occurs at external field strengths lower than $B_{c1}(T)$). The vortex phase itself occurs at external field strengths in between $B_{c1}(T)$ and $B_{c2}(T)$. 
 
\begin{figure}[H]
    \centering
    {\includegraphics[width=0.4\textwidth]{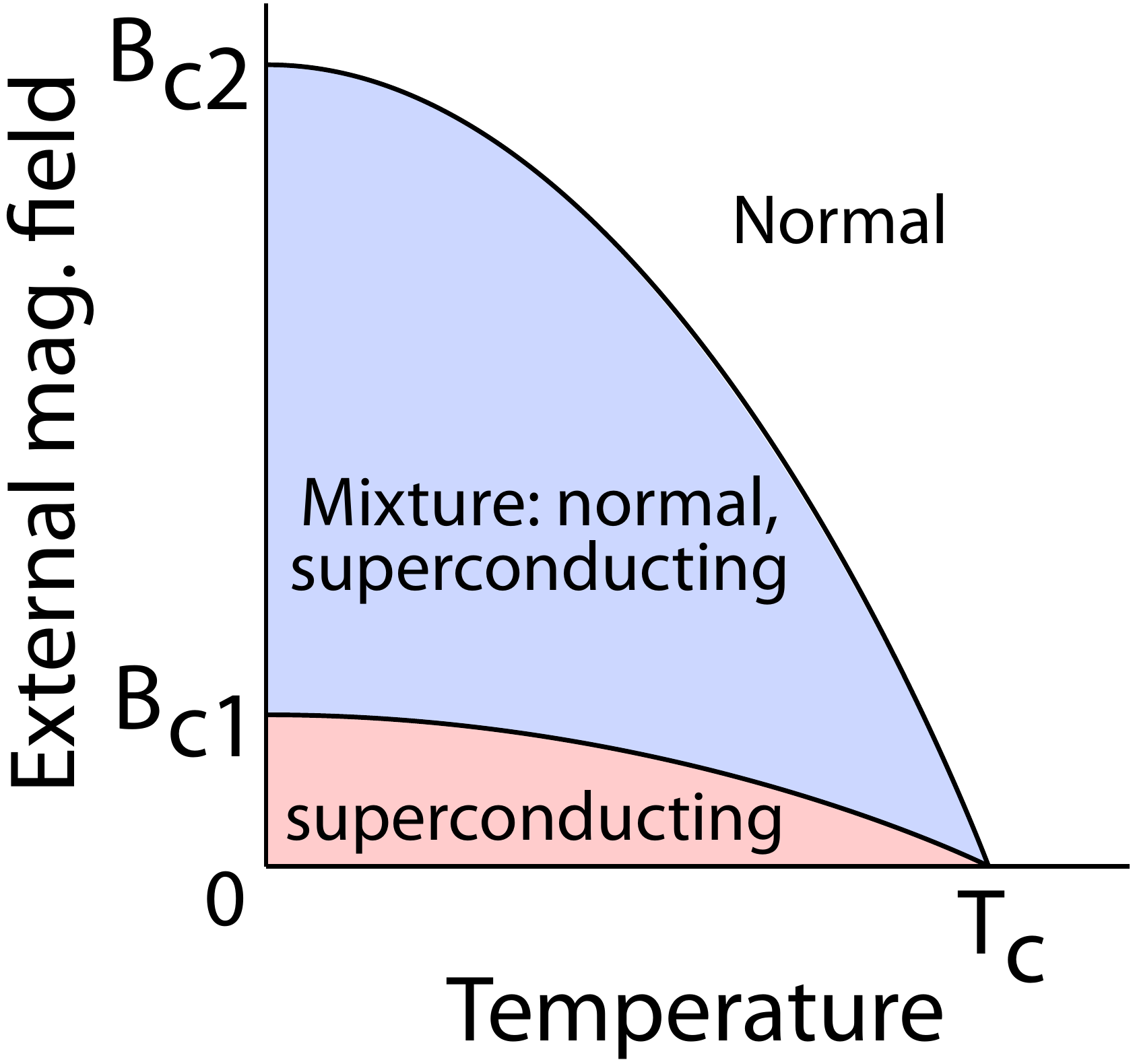}}
    \caption{Example diagram of the phase diagram of a type-II superconductor.}
    \label{fig:phase}
\end{figure}

Using a Gibbs free energy argument, we can show that \begin{equation}
    B_{c1}(T) = B_{c1}(0) \left( 1 - \left(\frac{T}{T_c}\right)^2 \right)
\end{equation} and that \begin{equation}
    B_{c2}(T) = B_{c2}(0) \left( 1 - \left(\frac{T}{T_c}\right)^2 \right)
\end{equation} \cite{Tilley}. These equations show that at $T = T_c$, $B_{c1}(T_c) = B_{c2}(T_c) = 0$. By the definition of the critical field strengths, $B_{c2}(T) > B_{c1}(T)$ for all $T < T_c$ for a type II superconductor (in which there is a non-trivial vortex phase region in a temperature versus external magnetic field phase diagram). 

Let us now consider how the external magnetic field impacts the magnetization of the superconducting sample and hence the magnetic field near the surface of the sample. First, let us consider the averaged magnetic field in a small region which nevertheless is bigger than the size of the vortices. In the Meissner phase, the magnetic field of the region is zero \cite{Josephson}. In the normal phase, the magnetic field is proportional to the external magnetic field, the constant of proportionality being determined from the magnetic permeability of the sample in its normal state \cite{Durrell}. Finally, in the vortex phase, the magnetization of the sample decays from a constant value to zero as the vortices shrink in size. Hence, at a fixed temperature, the magnetic field in the region varies with external magnetic field as follows: if we slowly increase the external magnetic field from zero, we will first see the field in the region flat at zero, then curve upwards with its first derivative increasing from zero to a fixed value, and finally the curve increasing linearly \cite{Fickett}. The transition from zero to curving upward occurs at the lower critical field strength, and the transition from curving upward to increasing linearly occurs at the upper critical field strength. 

Near a superconducting sample, the field will of course differ from the field inside the sample. Specifically the field near the sample will be higher than the field inside the sample \cite{Maxfield}. However, the field at a fixed small region near the sample will obey a similar pattern to the field in the inside region considered in the previous paragraph: the field will vary linearly with respect to the external magnetic field in the Meissner and normal phases, with the field curving upward in the vortex phase. The slope in the Meissner phase will be small (the field itself will be small but non-zero in the Meissner phase, with a value which depends linearly on the external magnetic field). A more precise discussion of these effects with reference to numerical simulations is provided in the appendices, but in short, we can qualitatively deduce the lower and upper critical field strengths from the graph of the field strength in a small region in the vicinity of the superconducting sample (i.e., the critical area in a Hall sensor), by noticing when the curve of the field as a function of the external field begins to curve upward and when the curve begins to straighten out into a constant linear line. 

\section{Experimental}

This section details our experimental procedure, discussing sample preparation, experimental design, and process.

\subsection{Sample Preparation}

For this experiment, we obtained a Vanadium disk, with a purity of $99.9\%$, and with a $24.26$ mm diameter and a $1.75$ mm thickness. Though oxygen and nitrogen impurities can affect the superconductive properties of the metal, experimental results have shown this noise is not statistically significant at this purity level \cite{Wexler}.

\subsection{Experimental Design}

For our experiment, we attached a brass vacuum chamber to the end of a cryostat probe. We conducted the experiment inside the vacuum chamber, placing the Vanadium disk in a specified temperature and external magnetic field and measuring the resulting magnetic field produced near the surface of the Vanadium disk. To produce the external magnetic field, we wound twelve layers of niobium tin superconducting wire around the vacuum chamber, with around one hundred and twenty turns per layer. We held the solenoid wires together by applying epoxy. Based on these numbers, we calculated the solenoid coefficient to be 0.045 T/A. We then calibrated our superconducting magnet power supply \cite{625} with this calculated coil constant. We have attached a more comprehensive review of the design and construction of the solenoid in the appendices. Inside the vacuum chamber, we suspended a copper block from the top of the vacuum chamber using screws. On opposite sides of the copper block, we embedded a Cernox sensor and a resistor to heat up the block, with both components being interfaced by a PID temperature controller \cite{32B}. On top of the block, we attached the Vanadium sample, and a PCB platform which contained our Hall bar interfaced by a Lock-in Amplifier \cite{SR830}. The Hall sensor which we used is a two dimensional indium arsenide Hall sensor which has a Hall coefficient of $12.5 \frac{\mu V}{T}$. 

When we run our experiment, we first submerge an outer layer of the cryostat in liquid nitrogen. Then, we fill the inner layer of the cryostat with liquid helium, submerging the probe. This sets the temperature of the sample to that of the boiling point of helium, 4.2K. We can then conduct measurements at 4.2K or higher using the resistor attached to the copper block. Then, we pump out gas from the liquid helium space to lower the temperature of the Vanadium sample, using the Cryocon's PID system to then stabilize our sample at pre-determined temperatures. Finally, we ramp the external magnetic field of our solenoid over a symmetric interval, measuring the detected magnetic field at discrete steps during the ramping process. We can repeat this pumping, stabilizing, and ramping process to conduct multiple measurements at temperatures lower than 4.2K. 

\subsection{Experimental Process}

In this section, we briefly review our experimental process. Note that we performed each of the cool downs discussed below before our final run of the experiment, to check and make sure that our electrical connections worked and that the experimental process worked. Below, we discuss each step of the experimental process as it is intended to be done when the experiment works. 

\subsubsection{Liquid Nitrogen Cool Down} 
To carry out the liquid nitrogen cool down, as well as the liquid helium cool down, we followed the procedures posted in Varian 4th floor lab. On our first liquid nitrogen cool down, we first checked for signs of vacuum leaks in every chamber. We noticed that there were leaks in the sealant where the magnet power leads pass through a hole in the probe stick lid. So, we mixed 5-minute epoxy and sealed up the vacuum leaks, which solved the problem.

After successfully adding liquid nitrogen to the cryostat, we used the Cryocon 32B temperature controller and the Cernox thermometer inside the probe to record the temperature of the sample every thirty seconds. We created a plot of the temperature of the sample over time that actively updated as the sample cooled, and we watched the temperature flatten off at approximately 78 K after approximately 90 minutes. When the sample reached equilibrium with the liquid nitrogen, we checked the electrical connection of all wires in the cryostat.

\subsubsection{Liquid Helium Cool Down Part One}

We inserted the transfer tube for liquid helium through an opening in the cryostat which had no blockages when pushing the tube down to the bottom of the cryostat. In particular, we ensured that the transfer tube did not hit our probe or the probe stick. We measured that the transfer tube could be inserted a maximum of 43 inches into the cryostat without hitting the probe, probe stick, or bottom of the dewar. Through transferring liquid helium we successfully cooled our sample to approximately 4.2 K. Then, we set the temperature of the Vanadium sample to five temperatures at or above 4.2K using our cryocon temperature control system: 4.29K, 4.38K, 4.46K, 4.63K, and 4.95K. At each temperature, we set the external magnetic field to a set of one hundred and one evenly spaced magnetic fields between -0.1T and 0.1T and measured the resulting magnetic field in the hall bar. 

\subsubsection{Liquid Helium Cool Down Part Two}

We repeated the above procedure for a liquid helium cool down. Then, we reduced the temperature of the sample below 4.2 K by pumping on the liquid helium space. We turned the pump knob fourth-fifths of a turn and watched the resulting pressure in the liquid helium space reduce. We noted the pressure dropping below atmospheric pressure, and the temperature quickly dropped as the pressure dropped. The data (pressure, temperature) is as follows. 500 Torr, 4.189K; 460 Torr, 3.77 K; 455 Torr, 3.754 K. We kept the knob of the pump that pumps on the liquid helium at four-fifths of a turn, and the pressure stabilized, as did the temperature. This test showed us that we can keep a stable temperature below 4.2 K by keeping the pump knob at a fixed position. 

After successfully reducing the temperature to 3.754 K, we slowly increased the pressure back to 760 Torr. We noticed that the temperature of the sample increased (as expected), but the temperature increased much slower than expected. Since the temperature of liquid helium is a function of the pressure in the space of the liquid helium, we would expect that the temperature would increase quickly as the pressure increased, which was not the case. While we are not sure what caused this delayed temperature rise, we speculate there could have been a liquid helium leak into the sample space that could have been the cause. If liquid helium had entered into the sample space, then since we were pumping on the sample space, the liquid helium in the sample space would keep the sample below 4.2 K.

We then pumped on the liquid helium space again by turning the pump knob four-fifths of a turn, then set the PID to a constant temperature (a value below 4.2 K), and we continued our data collection process for multiple values of temperature. For these lower temperatures we used a wider range of external magnetic fields: -0.16T to 0.16T for 3.3K, and -0.2T to 0.2T for 2.6K and 3.0K.

\section{Results}

\begin{figure}
    \centering
    {\includegraphics[width=0.5\textwidth]{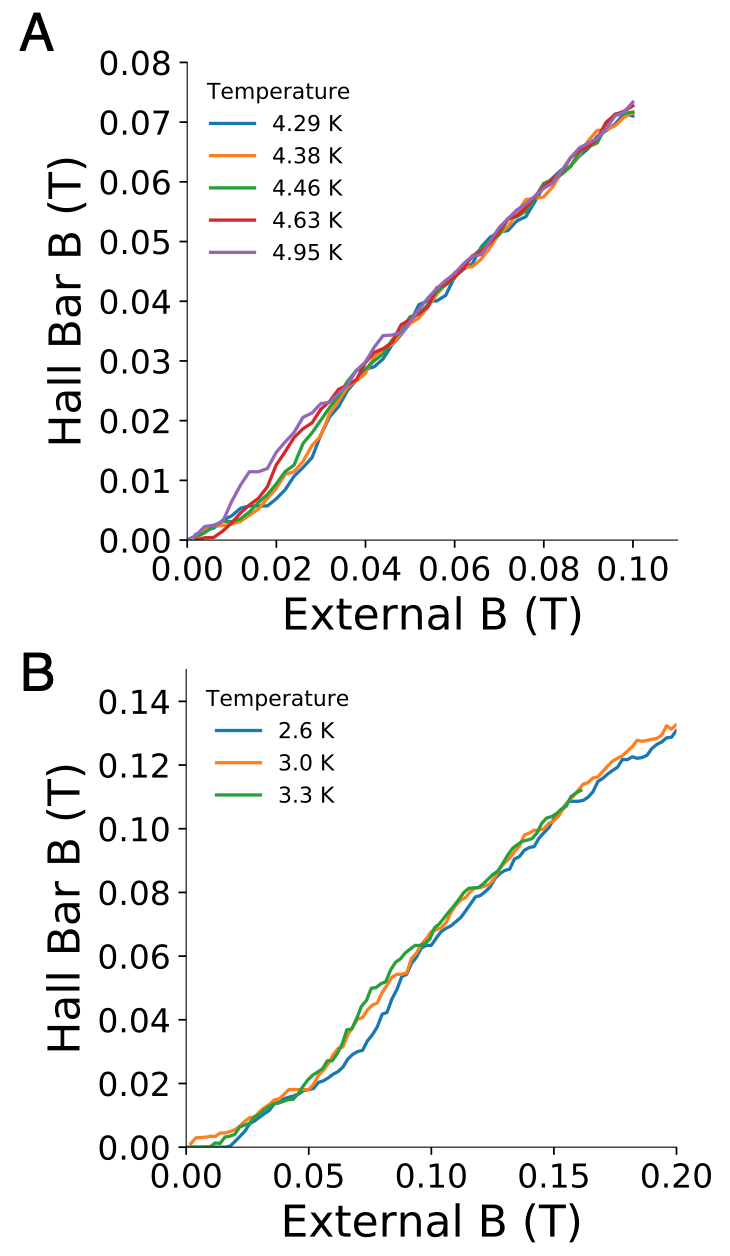}}
    \caption{Experimental data of the hall bar magnetic field as a function of the external magnetic field for temperatures at and above $4.29 K$ (A) and for temperatures at and below $3.3 K$ (B).}
    \label{fig:results}
\end{figure}

We varied both the temperature of the superconductor and the external magnetic field in the region of the sample, and we measured the magnetic field just above the surface of the Vanadium sample (Fig. \ref{fig:results}). To create these plots, note that the Hall bar outputs a voltage signal, which we then anti-symmetrized then multiplied by the hall coefficient to deduce the magnetic field at the Hall bar. Further details on how the hall voltage is processed to obtain the magnetic field being observed by the hall bar is contained in the appendices. 

\section{Analysis}

Let us denote the external magnetic field as $B_{ext}$ and the magnetic field detected by the hall bar to be $B_{hall}$. For each temperature, in the ideal case we expect to see a shallow line near the origin (at $0 \leq B_{ext} \leq B_{c1}$), then a curve connecting this shallow line to another steeper line $B_{c1} \leq B_{ext} \leq B_{c2}$, and then a steeper line heading off to the right at $B > B_{c2}$. These three regions correspond to the Meissner phase, Vortex phase, and Normal phase respectively. 

As a brief side-note before diving in to our estimates for the critical field values at different temperatures, we observed ridges spaced throughout our hall sensor plots. These ridges probably arise from the quantum hall effect, which would be operative in our hall sensor at temperatures around the boiling point of liquid helium. Also, it appears that there is an anomalous flat line near $B = 0T$ in the $2.6K$ plot, which may be due to the quantum hall effect or to some other unaccounted for source of noise. 

Accounting for the ridges commented on above, we can deduce the lower critical fields by identifying an initial shallow line region stemming from the origin. We can thus identify the following values for the lower critical field strengths: $0.074T$ for $2.9K$, $0.050T$ for $3.0K$, $0.044T$ for $3.3K$, $0.018T$ for $4.29K$, $0.0096T$ for $4.38K$, $0.0096T$ for $4.46K$, and $0.00595T$ for $4.63K$. Fitting these points in a quadratic regression of the form $B_{c1}(T) = a(1 - \left(\frac{T}{h}\right)^2)$, where $a = B_{c1}(0)$ and $h = T_c$, we get that $T_c = 4.70K$ and $B_{c1}(0) = 0.095T$. This implies that the data-point at $4.95K$ is above the critical temperature, which matches our observation that there is no phase transition apparent at $4.95K$. 

To calculate the upper field strengths, we will use the following technique. Let us fix a temperature and consider a plot corresponding to that temperature. We want to identify a bunch of points we know are in the normal phase, so that we can then compute a linear regression line corresponding to the normal phase. Then, we can identify $B_{c2}$ by observing when the graph begins to depart from the above regression line corresponding to the normal phase. 

In choosing which points we use to compute the normal phase regression line, we have two competing considerations: we want to choose more points to get a more accurate regression line, but we want to choose points which have a magnetic field higher than a high cutoff so that we are sure that these points are in fact in the normal phase of Vanadium and not in the vortex phase. 

From graphs of the field strengths in the literature \cite{Sekula, Wexler}, we know that for $T = 2.6K$, the upper critical field strength is below $0.18T$. Hence, we can compute a linear regression for points with magnetic fields from $0.18T$ to $0.2T$. Similarly, for $T = 3.0K$, the upper critical field strength is below $0.16T$, so we chose this as our cutoff for $T = 3.0K$. For $T = 3.3K$, we found that $0.14T$ was a good cutoff which was above the upper critical field for this temperature. Finally, for temperatures in the range $4K < T < 5K$, the upper critical field strength is bellow $0.08T$. Hence, we can compute a linear regression for points with magnetic fields from $0.08T$ to $0.1T$. We have attached in an appendix images of these normal phase regression lines superimposed on scatter plots of the original data points for the reader's reference. 

In summary, for $2.6K$ the upper critical field strength is $0.156T$, for $3.0K$ the upper critical field strength is $0.158$, for $3.3K$ the upper critical field strength is $0.0755T$, for $4.29K$ the upper critical field strength is $0.042T$, for $4.38K$ the upper critical field strength is $0.042T$, for $4.46K$ the upper critical field strength is $0.036T$, and for $4.63K$ the upper critical field strength the upper critical field strength is $0.024T$. Fitting these points in a quadratic regression of the form $B_{c2}(T) = a(1 - \left(\frac{T}{h}\right)^2)$, where $a = B_{c2}(0)$ and $h = T_c$, we get that $T_c = 4.85K$ and $B_{c1}(0) = 0.095T$. Again, this implies that the data-point at $4.95K$ is above the critical temperature. 

As a side note, the data for temperatures less than $4.2K$ is quite noisy, as the above graph of the data points and the regression curve illustrate. This can be attributed in part to temperature fluctuations: while for $T > 4K$ we were to stabilize the temperature to within $\pm 0.001K$, for temperatures less than $4K$ we saw temperature variations of $\pm 0.1K$ (and for $T = 3K$ we saw temperature variations of $\pm 0.2K$). These temperature fluctuations introduce error into our estimations of the upper critical field strengths, as will be seen shortly. 

However, we can be fairly confident that we have a good estimate for the critical temperature, since our lower critical field estimates suggest $T_c = 4.70K$ and our upper critical field estimates suggest $T_c = 4.85K$. We can take a rough average of these temperatures to get $T_c = 4.8K$ as our estimate of the critical temperature of our Vanadium sample. Using this as an estimate, we can re-compute regression curves for the lower and upper critical field strengths of the form $B_{c1}(T) = a_1(1 - \left(\frac{T}{T_c}\right)^2)$ and $B_{c2}(T) = a_2(1 - \left(\frac{T}{T_c}\right)^2)$, where $T_c = 4.8K$. Doing this, we get that $a_1 = B_{c1}(0) = 0.091T$ and $a_2 = B_{c2}(0) = 0.216T$. Using this, we produced a phase diagram for our Vanadium sample (Figure 2).

\begin{figure}
    \centering
    \includegraphics[scale=0.8]{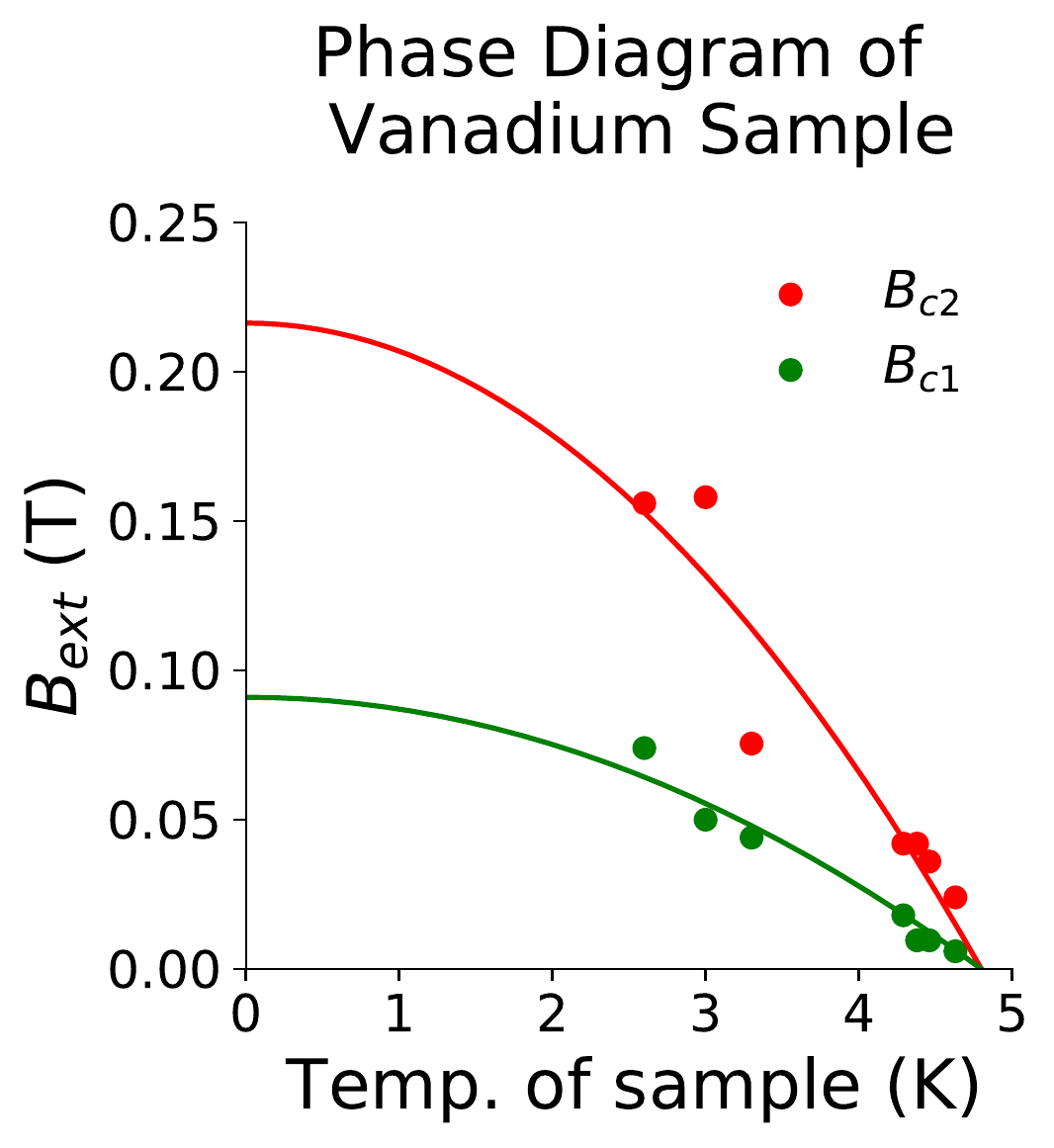}
    \caption{Plot of the upper critical field strength ($B_{c2}$) and lower critical field strength ($B_{c1}$) as a function of temperature. The region below the green line corresponds to the Meissner phase, the region between the green and red lines corresponds to the Vortex phase, and the region above the red line corresponds to the normal phase.}
    \label{fig:phase_diagram}
\end{figure}

\section{Conclusion}

Through our hall sensor based measurement method, we were able to detect our Vanadium sample in three superconducting phases and to produce estimates of the upper and lower critical field strength curves for our Vanadium sample. Our estimates were lower than those measured by Sekula and Kernohan, though they are within the same ball park \cite{Sekula}. Specifically, we measured a $T_c$ of $4.8K$ while Sekula and Kernohan measured a $T_c$ of $5.43K$ for pure Vanadium, and we measured $B_{c1}(0) = 0.091T$ and $B_{c2}(0) = 0.216T$ while Sekula and Kernohan measured that $B_{c1}(0) = 0.14T$ and $B_{c2}(0) \approx 0.27T$. It is plausible that the impurities in our Vanadium sample caused this discrepancy: we used a $99.9\%$ pure Vanadium sample, whose impurities were not detailed by the distributor who we bought the sample from. These impurities could interfere with the superconducting properties of Vanadium, changing the resulting critical temperature and critical field measurements. 

For future work, we would like to find ways of improving the precision of our hall sensor based method for measuring the critical fields of a superconducting sample. Specifically, the hall sensor graphs were quite noisy, especially with ridges caused from the quantum hall effect, which made it particularly difficult to distinguish the vortex phase from the normal phase. Some possible future improvements could be to sample substantially more data points, use a more precise hall bar or a different magnetometer system which does not suffer from distortions due to the quantum hall effect, and to take repeated measurements at the same temperature and average these measurements out. The latter improvement could allow us to eliminate fluctuations due to noise in each phase, thereby producing a cleaner plot for each temperature, with a shallow straight line in the Meissner phase, a curve in the vortex phase, and a steeper straight line in the normal phase. 

\section{Author declarations}
The authors have no conflicts to disclose.
\begin{acknowledgments}
We wish to acknowledge the support and help which Prof. David Goldhaber-Gordon, Rick Pam, Joe Finney, and Kaan Alp Yay provided for us in coming up with and physically implementing this experiment. We further wish to acknowledge Matthew Chuck and Mehmet Solyani in the machine shop for the Stanford physics department, who machined crucial parts of our experimental apparatus, and Connie Hsueh from the Goldhaber-Gordon Group, who helped wire bond our Hall bar to our PCB. The three authors designed and carried out the experiment for an undergraduate experimental physics lab, and are listed in alphabetical order.

\end{acknowledgments}

\nocite{*}

\bibliography{vanadium}

\providecommand{\noopsort}[1]{}\providecommand{\singleletter}[1]{#1}%
\begin{thebibliography}{18}%
\makeatletter
\providecommand \@ifxundefined [1]{%
 \@ifx{#1\undefined}
}%
\providecommand \@ifnum [1]{%
 \ifnum #1\expandafter \@firstoftwo
 \else \expandafter \@secondoftwo
 \fi
}%
\providecommand \@ifx [1]{%
 \ifx #1\expandafter \@firstoftwo
 \else \expandafter \@secondoftwo
 \fi
}%
\providecommand \natexlab [1]{#1}%
\providecommand \enquote  [1]{``#1''}%
\providecommand \bibnamefont  [1]{#1}%
\providecommand \bibfnamefont [1]{#1}%
\providecommand \citenamefont [1]{#1}%
\providecommand \href@noop [0]{\@secondoftwo}%
\providecommand \href [0]{\begingroup \@sanitize@url \@href}%
\providecommand \@href[1]{\@@startlink{#1}\@@href}%
\providecommand \@@href[1]{\endgroup#1\@@endlink}%
\providecommand \@sanitize@url [0]{\catcode `\\12\catcode `\$12\catcode
  `\&12\catcode `\#12\catcode `\^12\catcode `\_12\catcode `\%12\relax}%
\providecommand \@@startlink[1]{}%
\providecommand \@@endlink[0]{}%
\providecommand \url  [0]{\begingroup\@sanitize@url \@url }%
\providecommand \@url [1]{\endgroup\@href {#1}{\urlprefix }}%
\providecommand \urlprefix  [0]{URL }%
\providecommand \Eprint [0]{\href }%
\providecommand \doibase [0]{https://doi.org/}%
\providecommand \selectlanguage [0]{\@gobble}%
\providecommand \bibinfo  [0]{\@secondoftwo}%
\providecommand \bibfield  [0]{\@secondoftwo}%
\providecommand \translation [1]{[#1]}%
\providecommand \BibitemOpen [0]{}%
\providecommand \bibitemStop [0]{}%
\providecommand \bibitemNoStop [0]{.\EOS\space}%
\providecommand \EOS [0]{\spacefactor3000\relax}%
\providecommand \BibitemShut  [1]{\csname bibitem#1\endcsname}%
\let\auto@bib@innerbib\@empty
\bibitem [{\citenamefont {Leplae}\ \emph {et~al.}(1972)\citenamefont {Leplae},
  \citenamefont {Mancini},\ and\ \citenamefont {Umezawa}}]{Leplae}%
  \BibitemOpen
  \bibfield  {author} {\bibinfo {author} {\bibfnamefont {L.}~\bibnamefont
  {Leplae}}, \bibinfo {author} {\bibfnamefont {F.}~\bibnamefont {Mancini}},\
  and\ \bibinfo {author} {\bibfnamefont {H.}~\bibnamefont {Umezawa}},\
  }\bibfield  {title} {\bibinfo {title} {Magnetic properties of vanadium and
  niobium at $t=0\ifmmode^\circ\else\textdegree\fi{}$k},\ }\href
  {https://doi.org/10.1103/PhysRevB.6.4178} {\bibfield  {journal} {\bibinfo
  {journal} {Phys. Rev. B}\ }\textbf {\bibinfo {volume} {6}},\ \bibinfo {pages}
  {4178} (\bibinfo {year} {1972})}\BibitemShut {NoStop}%
\bibitem [{\citenamefont {Tilley}\ and\ \citenamefont {Tilley}(1990)}]{Tilley}%
  \BibitemOpen
  \bibfield  {author} {\bibinfo {author} {\bibfnamefont {D.~R.}\ \bibnamefont
  {Tilley}}\ and\ \bibinfo {author} {\bibfnamefont {J.}~\bibnamefont
  {Tilley}},\ }\href@noop {} {\emph {\bibinfo {title} {Superfluidity and
  Superconductivity}}}\ (\bibinfo  {publisher} {Graduate Student Series in
  Physics},\ \bibinfo {year} {1990})\BibitemShut {NoStop}%
\bibitem [{\citenamefont {Radebaugh}\ and\ \citenamefont
  {Keesom}(1966)}]{Radebaugh}%
  \BibitemOpen
  \bibfield  {author} {\bibinfo {author} {\bibfnamefont {R.}~\bibnamefont
  {Radebaugh}}\ and\ \bibinfo {author} {\bibfnamefont {P.~H.}\ \bibnamefont
  {Keesom}},\ }\bibfield  {title} {\bibinfo {title} {Low-temperature
  thermodynamic properties of vanadium. ii. mixed state},\ }\href
  {https://doi.org/10.1103/PhysRev.149.217} {\bibfield  {journal} {\bibinfo
  {journal} {Phys. Rev.}\ }\textbf {\bibinfo {volume} {149}},\ \bibinfo {pages}
  {217} (\bibinfo {year} {1966})}\BibitemShut {NoStop}%
\bibitem [{\citenamefont {Beebe}\ \emph {et~al.}(2017)\citenamefont {Beebe},
  \citenamefont {Valente-Feliciano}, \citenamefont {Beringer}, \citenamefont
  {Creeden}, \citenamefont {Madaras}, \citenamefont {Li}, \citenamefont {Yang},
  \citenamefont {Phillips}, \citenamefont {Reece},\ and\ \citenamefont
  {Lukaszew}}]{Beebe}%
  \BibitemOpen
  \bibfield  {author} {\bibinfo {author} {\bibfnamefont {M.~R.}\ \bibnamefont
  {Beebe}}, \bibinfo {author} {\bibfnamefont {A.-M.}\ \bibnamefont
  {Valente-Feliciano}}, \bibinfo {author} {\bibfnamefont {D.~B.}\ \bibnamefont
  {Beringer}}, \bibinfo {author} {\bibfnamefont {J.~A.}\ \bibnamefont
  {Creeden}}, \bibinfo {author} {\bibfnamefont {S.~E.}\ \bibnamefont
  {Madaras}}, \bibinfo {author} {\bibfnamefont {Z.}~\bibnamefont {Li}},
  \bibinfo {author} {\bibfnamefont {K.}~\bibnamefont {Yang}}, \bibinfo {author}
  {\bibfnamefont {L.}~\bibnamefont {Phillips}}, \bibinfo {author}
  {\bibfnamefont {C.~E.}\ \bibnamefont {Reece}},\ and\ \bibinfo {author}
  {\bibfnamefont {R.~A.}\ \bibnamefont {Lukaszew}},\ }\bibfield  {title}
  {\bibinfo {title} {Temperature and microstructural effects on the
  superconducting properties of niobium thin films},\ }\href
  {https://doi.org/10.1109/TASC.2016.2632420} {\bibfield  {journal} {\bibinfo
  {journal} {IEEE Transactions on Applied Superconductivity}\ }\textbf
  {\bibinfo {volume} {27}},\ \bibinfo {pages} {1} (\bibinfo {year}
  {2017})}\BibitemShut {NoStop}%
\bibitem [{\citenamefont {Sekula}\ and\ \citenamefont
  {Kernohan}(1972)}]{Sekula}%
  \BibitemOpen
  \bibfield  {author} {\bibinfo {author} {\bibfnamefont {S.~T.}\ \bibnamefont
  {Sekula}}\ and\ \bibinfo {author} {\bibfnamefont {R.~H.}\ \bibnamefont
  {Kernohan}},\ }\bibfield  {title} {\bibinfo {title} {Magnetic properties of
  superconducting vanadium},\ }\href {https://doi.org/10.1103/PhysRevB.5.904}
  {\bibfield  {journal} {\bibinfo  {journal} {Phys. Rev. B}\ }\textbf {\bibinfo
  {volume} {5}},\ \bibinfo {pages} {904} (\bibinfo {year} {1972})}\BibitemShut
  {NoStop}%
\bibitem [{\citenamefont {Wexler}\ and\ \citenamefont {Corak}(1952)}]{Wexler}%
  \BibitemOpen
  \bibfield  {author} {\bibinfo {author} {\bibfnamefont {A.}~\bibnamefont
  {Wexler}}\ and\ \bibinfo {author} {\bibfnamefont {W.~S.}\ \bibnamefont
  {Corak}},\ }\bibfield  {title} {\bibinfo {title} {Superconductivity of
  vanadium},\ }\href {https://doi.org/10.1103/PhysRev.85.85} {\bibfield
  {journal} {\bibinfo  {journal} {Phys. Rev.}\ }\textbf {\bibinfo {volume}
  {85}},\ \bibinfo {pages} {85} (\bibinfo {year} {1952})}\BibitemShut {NoStop}%
\bibitem [{\citenamefont {Brown}\ \emph {et~al.}(1953)\citenamefont {Brown},
  \citenamefont {Zemansky},\ and\ \citenamefont {Boorse}}]{Brown}%
  \BibitemOpen
  \bibfield  {author} {\bibinfo {author} {\bibfnamefont {A.}~\bibnamefont
  {Brown}}, \bibinfo {author} {\bibfnamefont {M.~W.}\ \bibnamefont
  {Zemansky}},\ and\ \bibinfo {author} {\bibfnamefont {H.~A.}\ \bibnamefont
  {Boorse}},\ }\bibfield  {title} {\bibinfo {title} {The superconducting and
  normal heat capacities of niobium},\ }\href
  {https://doi.org/10.1103/PhysRev.92.52} {\bibfield  {journal} {\bibinfo
  {journal} {Phys. Rev.}\ }\textbf {\bibinfo {volume} {92}},\ \bibinfo {pages}
  {52} (\bibinfo {year} {1953})}\BibitemShut {NoStop}%
\bibitem [{\citenamefont {Ohuchi}\ \emph {et~al.}(2022)\citenamefont {Ohuchi},
  \citenamefont {Ueki},\ and\ \citenamefont {Kita}}]{Ohuchi}%
  \BibitemOpen
  \bibfield  {author} {\bibinfo {author} {\bibfnamefont {M.}~\bibnamefont
  {Ohuchi}}, \bibinfo {author} {\bibfnamefont {H.}~\bibnamefont {Ueki}},\ and\
  \bibinfo {author} {\bibfnamefont {T.}~\bibnamefont {Kita}},\ }\bibfield
  {title} {\bibinfo {title} {Charging in the vortex lattice of type-{II}
  superconductors},\ }\bibfield  {journal} {\bibinfo  {journal} {Physical
  Review B}\ }\textbf {\bibinfo {volume} {105}},\ \href
  {https://doi.org/10.1103/physrevb.105.064514} {10.1103/physrevb.105.064514}
  (\bibinfo {year} {2022})\BibitemShut {NoStop}%
\bibitem [{\citenamefont {Zaytseva}\ \emph {et~al.}(2020)\citenamefont
  {Zaytseva}, \citenamefont {Abaloszew}, \citenamefont {Camargo}, \citenamefont
  {Syryanyy},\ and\ \citenamefont {Cieplak}}]{Iryna}%
  \BibitemOpen
  \bibfield  {author} {\bibinfo {author} {\bibfnamefont {I.}~\bibnamefont
  {Zaytseva}}, \bibinfo {author} {\bibfnamefont {A.}~\bibnamefont {Abaloszew}},
  \bibinfo {author} {\bibfnamefont {B.~C.}\ \bibnamefont {Camargo}}, \bibinfo
  {author} {\bibfnamefont {Y.}~\bibnamefont {Syryanyy}},\ and\ \bibinfo
  {author} {\bibfnamefont {M.~Z.}\ \bibnamefont {Cieplak}},\ }\bibfield
  {title} {\bibinfo {title} {Upper critical field and superconductor-metal
  transition in ultrathin niobium films},\ }\bibfield  {journal} {\bibinfo
  {journal} {Sci Rep}\ }\textbf {\bibinfo {volume} {10}},\ \href
  {https://doi.org/10.1038/s41598-020-75968-9} {10.1038/s41598-020-75968-9}
  (\bibinfo {year} {2020})\BibitemShut {NoStop}%
\bibitem [{\citenamefont {Ma}\ \emph {et~al.}(2021)\citenamefont {Ma},
  \citenamefont {Gornicka}, \citenamefont {Lefèvre}, \citenamefont {Yang},
  \citenamefont {Rønnow}, \citenamefont {Jeschke}, \citenamefont {Klimczuk},\
  and\ \citenamefont {von Rohr}}]{Ma}%
  \BibitemOpen
  \bibfield  {author} {\bibinfo {author} {\bibfnamefont {K.}~\bibnamefont
  {Ma}}, \bibinfo {author} {\bibfnamefont {K.}~\bibnamefont {Gornicka}},
  \bibinfo {author} {\bibfnamefont {R.}~\bibnamefont {Lefèvre}}, \bibinfo
  {author} {\bibfnamefont {Y.}~\bibnamefont {Yang}}, \bibinfo {author}
  {\bibfnamefont {H.~M.}\ \bibnamefont {Rønnow}}, \bibinfo {author}
  {\bibfnamefont {H.~O.}\ \bibnamefont {Jeschke}}, \bibinfo {author}
  {\bibfnamefont {T.}~\bibnamefont {Klimczuk}},\ and\ \bibinfo {author}
  {\bibfnamefont {F.~O.}\ \bibnamefont {von Rohr}},\ }\bibfield  {title}
  {\bibinfo {title} {Superconductivity with high upper critical field in the
  cubic centrosymmetric η-carbide nb4rh2c1−δ},\ }\href
  {https://doi.org/10.1021/acsmaterialsau.1c00011} {\bibfield  {journal}
  {\bibinfo  {journal} {ACS Materials Au}\ }\textbf {\bibinfo {volume} {1}},\
  \bibinfo {pages} {55} (\bibinfo {year} {2021})},\ \Eprint
  {https://arxiv.org/abs/https://doi.org/10.1021/acsmaterialsau.1c00011}
  {https://doi.org/10.1021/acsmaterialsau.1c00011} \BibitemShut {NoStop}%
\bibitem [{Jos(2017)}]{Josephson}%
  \BibitemOpen
  \bibfield  {title} {\bibinfo {title} {Superconductivity: The meissner effect,
  persistent currents and the josephson lab guide}\ }(\bibinfo {year}
  {2017})\BibitemShut {NoStop}%
\bibitem [{\citenamefont {Durrell}\ \emph {et~al.}(2014)\citenamefont
  {Durrell}, \citenamefont {Dennis}, \citenamefont {Jaroszynski}, \citenamefont
  {Ainslie}, \citenamefont {Palmer}, \citenamefont {Shi}, \citenamefont
  {Campbell}, \citenamefont {Hull}, \citenamefont {Strasik}, \citenamefont
  {Hellstrom},\ and\ \citenamefont {Cardwell}}]{Durrell}%
  \BibitemOpen
  \bibfield  {author} {\bibinfo {author} {\bibfnamefont {J.~H.}\ \bibnamefont
  {Durrell}}, \bibinfo {author} {\bibfnamefont {A.~R.}\ \bibnamefont {Dennis}},
  \bibinfo {author} {\bibfnamefont {J.}~\bibnamefont {Jaroszynski}}, \bibinfo
  {author} {\bibfnamefont {M.~D.}\ \bibnamefont {Ainslie}}, \bibinfo {author}
  {\bibfnamefont {K.~G.~B.}\ \bibnamefont {Palmer}}, \bibinfo {author}
  {\bibfnamefont {Y.-H.}\ \bibnamefont {Shi}}, \bibinfo {author} {\bibfnamefont
  {A.~M.}\ \bibnamefont {Campbell}}, \bibinfo {author} {\bibfnamefont
  {J.}~\bibnamefont {Hull}}, \bibinfo {author} {\bibfnamefont {M.}~\bibnamefont
  {Strasik}}, \bibinfo {author} {\bibfnamefont {E.~E.}\ \bibnamefont
  {Hellstrom}},\ and\ \bibinfo {author} {\bibfnamefont {D.~A.}\ \bibnamefont
  {Cardwell}},\ }\bibfield  {title} {\bibinfo {title} {A trapped field of 17.6
  t in melt-processed, bulk gd-ba-cu-o reinforced with shrink-fit steel},\
  }\href {https://doi.org/10.1088/0953-2048/27/8/082001} {\bibfield  {journal}
  {\bibinfo  {journal} {Superconductor Science and Technology}\ }\textbf
  {\bibinfo {volume} {27}},\ \bibinfo {pages} {082001} (\bibinfo {year}
  {2014})}\BibitemShut {NoStop}%
\bibitem [{\citenamefont {Fickett}(1985)}]{Fickett}%
  \BibitemOpen
  \bibfield  {author} {\bibinfo {author} {\bibfnamefont {F.~R.}\ \bibnamefont
  {Fickett}},\ }\bibfield  {title} {\bibinfo {title} {Standards for measurement
  of the critical fields of superconductors},\ }\bibfield  {journal} {\bibinfo
  {journal} {J. Res. Natl. Bur. Stand. (U.S.); (United States)}\ }\textbf
  {\bibinfo {volume} {90}},\ \href {https://doi.org/10.6028/jres.090.007}
  {10.6028/jres.090.007} (\bibinfo {year} {1985})\BibitemShut {NoStop}%
\bibitem [{\citenamefont {Maxfield}\ and\ \citenamefont
  {McLean}(1965)}]{Maxfield}%
  \BibitemOpen
  \bibfield  {author} {\bibinfo {author} {\bibfnamefont {B.~W.}\ \bibnamefont
  {Maxfield}}\ and\ \bibinfo {author} {\bibfnamefont {W.~L.}\ \bibnamefont
  {McLean}},\ }\bibfield  {title} {\bibinfo {title} {Superconducting
  penetration depth of niobium},\ }\href
  {https://doi.org/10.1103/PhysRev.139.A1515} {\bibfield  {journal} {\bibinfo
  {journal} {Phys. Rev.}\ }\textbf {\bibinfo {volume} {139}},\ \bibinfo {pages}
  {A1515} (\bibinfo {year} {1965})}\BibitemShut {NoStop}%
\bibitem [{625()}]{625}%
  \BibitemOpen
  \href@noop {} {\emph {\bibinfo {title} {User’s Manual Model 625
  Superconducting Magnet Power Supply}}}\BibitemShut {NoStop}%
\bibitem [{32B()}]{32B}%
  \BibitemOpen
  \href@noop {} {\emph {\bibinfo {title} {User's Guide Model 32 \& 32B
  Cryogenic Temperature Controller}}}\BibitemShut {NoStop}%
\bibitem [{SR8()}]{SR830}%
  \BibitemOpen
  \href@noop {} {\emph {\bibinfo {title} {MODEL SR830 DSP Lock-In
  Amplifier}}}\BibitemShut {NoStop}%
\bibitem [{\citenamefont {Ekin}(2006)}]{Ekin2006}%
  \BibitemOpen
  \bibfield  {author} {\bibinfo {author} {\bibfnamefont {J.}~\bibnamefont
  {Ekin}},\ }\href@noop {} {\emph {\bibinfo {title} {Experimental Techniques
  for Low-Temperature Measurements: Cryostat Design, Material Properties and
  Superconductor Critical-Current Testing}}}\ (\bibinfo  {publisher} {Oxford
  University Press},\ \bibinfo {year} {2006})\BibitemShut {NoStop}%
\end{thebibliography}%

\clearpage

\appendix
\counterwithin{figure}{section}

\section{Solenoid Design}

In this appendix, we will motivate the design for our solenoid as follows. First, we will review the derivation for formulas determining the number of turns per layer $N$, number of total layers $t$, and total length of wire desired based $l$ as a function of the thickness of the solenoid wires $t$, the radius of the vacuum chamber $R$, the height of the solenoid $h$, the maximum current we can use $I_{max}$, and the smallest upper bound for the magnetic field which we would like the solenoid to be able to produce $B_{max}$. Then, based on the number of layers we chose and the actual turns per layer we made in constructing the probe, we will determine the solenoid coefficient $k$ of the solenoid.

In terms of actual numbers, we had $t = 0.33 mm$, $R = 1.98 cm$, $h = 4.3 cm$ $I_{max} = 25A$, and $B_{max} = 1T$. We chose $I_{max} = 25A$ since this was the largest current we could secure, and $B_{max} = 1T$ since that value was well above the upper critical field strength of Vanadium at absolute zero. We ended up choosing $12$ layers, intending for $130$ turns per layer, but in the construction of our solenoid we ended up with $121$ layers per turn. Hence, while our intended solenoid coefficient $k$ was $0.045 T/A$, our actual solenoid coefficient was $0.0424 T/A$.

\subsection{Number of Layers and Number of Turns per Layer}

The thickness of the solenoid wires is $t$. Hence, if there are $N$ number of turns per layer, as the solenoid as a height $h$, we get that $Nt \leq h$. To maximize the number of turns per layer, and hence minimize the number of layers, we get that $\boxed{N(h, t) = \left\lfloor \frac{h}{t} \right\rfloor}$. For $h = 4.3 cm$ and $t = 0.33 mm$, $N = 130$ turns per layer. As stated before, in the actual construction of our solenoid, we in fact had $\boxed{N = 121}$ turns per layer due to gaps which crept in during the winding process. 

Now, let the solenoid have $l$ layers. Then, the number of turns per unit length equals $n = \frac{Nl}{Nt} = \frac{l}{t}$. Hence, the maximum magnetic field produced inside the solenoid equals $B = \mu_0 n I_{max} = \mu_0 I_{max} \frac{l}{t}$. We want this field to exceed $B_{max}$: $B_{max} \leq \mu_0 I_{max} \frac{l}{t}$, $l \geq \frac{B_{max}}{\mu_0 I_{max}} t$. To minimize $l$, we get that $l(B_{max}, I_{max}, t) = \frac{B_{max} t}{\mu_0 I_{max}}$. Note that we really want an even number of layers so that the wires of the solenoid enter and exit the vacuum chamber on the same side; this results in the following expression for the number of layers: $\boxed{l(B_{max}, I_{max}, t) = 2 \left\lceil \frac{1}{2} \frac{B_{max} t}{\mu_0 I_{max}} \right\rceil}$. With $t = 0.33 mm$, $B_{max} = 1T$, $I_{max} = 25A$, and $B_{max} = 1T$, we had $\boxed{l = 12}$ layers. 

\subsection{Total Length of Wire}

We have $l$ layers of wire, each layer having $N$ turns. For the $i$-th layer, the radius of a single turn equals $R + (i - \frac{1}{2}) t$. Hence, 

\begin{equation*}
\begin{split}
    L = \sum_{i = 1}^l N 2 \pi (R + (i - \frac{1}{2}) t) = 2 \pi N \left((R - \frac{t}{2}) l + \frac{l(l+1)}{2}t\right) \\ = 2 \pi N l \left(R + \frac{tl}{2}\right)
\end{split}
\end{equation*}

Now, in actually constructing the solenoid, we could empirically measure the total number of turns used, and hence we have a more accurate estimate of $N$ by diving this total number by the number of layers: $N = 1450/12$. Further, we have that $l = 12$, $R = 1.96 cm$, and $t = 0.33 mm$. Hence, 

\begin{equation*}
\begin{split}
    \boxed{L = 188 m}.
\end{split}
\end{equation*}

This length is well within the total amount of niobium tin wires which we had.

\subsection{External Magnetic Field Produced As A Function of Current}

The magnetic field inside a solenoid as a function of current in the wire is given by,

\begin{equation*}
    \boxed{B(I) = k I} 
\end{equation*}

where 

\begin{equation*}
    \boxed{k(t, B_{max}, I_{max}) = \mu_0 \frac{Nl}{h}}
\end{equation*}

With $Nl = 1450$ and $h = 4.3 cm$, we get that 

\begin{equation*}
    \boxed{k(t, B_{max}, I_{max}) = 0.0424 T/A}.
\end{equation*}

\section{Solenoid Construction}

For our solenoid, we used a niobium tin wire with a thickness of $0.33$ millimeters which was wrapped around a vacuum chamber with outer radius of $1.96$ centimeters across a height of around $4.3$ centimeters. As stated in the solenoid design section, we imposed that the total number of turns was $12$. We intended for there to be $130$ turns per layer, but in the process of winding the solenoid we found that we observed an estimated $1450/12 \approx 121$ turns per layer, which we hypothesize is due to gaps between the wires that we observed.  

\begin{figure}
    \centering
    \includegraphics[scale=0.6]{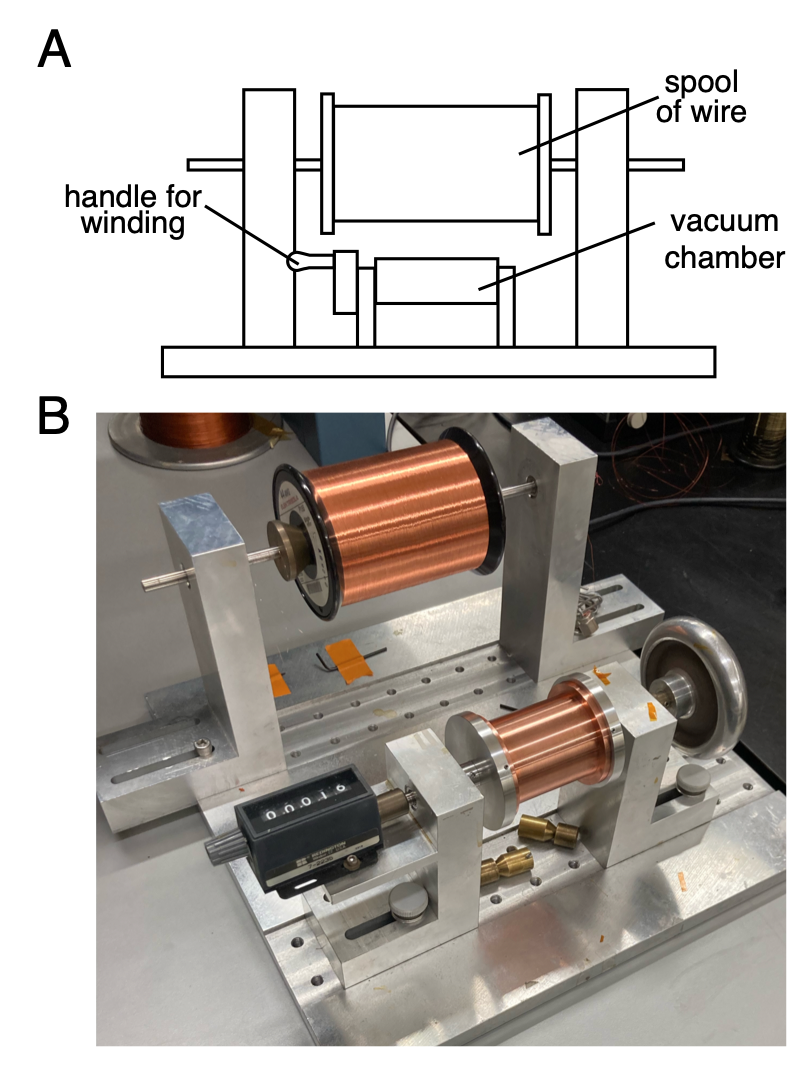}
    \caption{Diagram (A) and photo (B) of our solenoid winding machine.}
    \label{fig:solenoid_rig}
\end{figure}

The solenoid winding machine is a platform connected to three objects: a structure to hold and rotate a spool of wire, a structure connected to a handle for rotation and in the other side connected to a rod which rotates, and finally a structure connected to a rod which rotates and a mechanism which counts the number of turns which the rod rotates (Fig. \ref{fig:solenoid_rig}). To attach our vacuum chamber to this mechanism, we first constructed metal disks which on one end can fix itself to the rod associated with one of the latter two structures and which in the other end can snugly fit the plates on the top and the bottom of the vacuum chamber. A set screw was used to attach each plate of the vacuum chamber to the appropriate metal disk. Note that we should place the top of the vacuum chamber, with the holes, on the metal disk attached to the structure with the handle. 

After setting up the solenoid to the solenoid wire winding apparatus as described above, we wound the solenoid as follows. First, we placed the spool of wire in the first structure mentioned in the previous paragraph. Then, we placed the spool of wire in the position indicated in the diagram. Afterwards, we unrolled a few inches of wire then taped the wire down near the top of the chamber, i.e. the side of the vacuum chamber next to the handle. Then, we produced epoxy by mixing Stycast 2850ft with Catalyst 23lv with a weight ratio of $100:7.5$. After this, we began slowly winding the solenoid. While the solenoid is being wound around the vacuum, apply epoxy to the wires. We suggest that two people perform this step: one person does the winding, and the second person guides the wire, ensuring it is taut, and applies the epoxy. When a layer is wound, we begin winding the next layer by going on top of the previous layer and leading the solenoid wire in the other direction. 

After the above procedure is completed, all the solenoid wire layers should be wound on the vacuum chamber. The wire must then be taped to the top plate of the vacuum chamber, before some additional inches of wire are wound out. The wire is then cut from the spool. Afterwards, apply a thick layer of epoxy around the entire solenoid. We can then cut off the leading wires from the first and last layers, leaving an inch. We then placed tubing on each inch of leading wire. 

Finally, we rotated the solenoid at a slow and steady velocity for one hour. Ideally, this should be done overnight, i.e. by using some weak motor attached to one of the two rods attached to either end of the vacuum chamber. Doing it for an hour is fine, but will probably lead to some epoxy settling at the bottom (since the epoxy won't be fully dry yet). This extra epoxy at the bottom should be scraped off until the vacuum chamber can fit into the cryostat. 

\section{Probe Design}

In this section, we will go into more depth in our probe design. In particular, we wish to discuss three crucial elements of our experimental apparatus: a brass vacuum chamber, a bronze vacuum chamber lid, and a copper platform. Below, we describe and motivate our design choices for these three elements. 

\subsection{Vacuum Chamber}

The vacuum chamber had to fit inside of the dewar, but also have space to enclose our sample, Hall bar, Cernox sensor and connection wires. We require 8 wires stemming from a DB-9 connector at the top of the probe, broken down as follows: 2-terminal Cernox measurement, 2-terminal resistive heater control, 4-terminal Hall bar measurement. A more detailed description of electronics can be found in Appendix E. Additionally, we had to wind a solenoid around the vacuum chamber, so we needed flanges to hold the solenoid in place. With these design goals, we chose a cylindrical chamber with a flange at the top and bottom that allowed for a solenoid with a total thickness of 5 mm. We minimized the thickness of the walls of the vacuum chamber while still providing structural integrity by choosing a wall thickness of 2 mm. We placed a groove for an indium seal at the top of the vacuum chamber, as well as 12 screw holes for attaching the vacuum chamber to the lid using brass screws. We chose brass screws since they are non-magnetic. The thickness of the flanges is 6.35 mm to provide enough room for the screws to attach to the vacuum chamber lid.

\subsection{Vacuum Chamber Lid}

The brass vacuum chamber lid was designed to attach and be vacuum sealed to the probe stick, to be attached to the vacuum chamber with screws, and to secure the sample in the center of the solenoid. To achieve the first of these goals, we added a protruding cylindrical piece that fit snugly around the probe stick; the outer diameter of the probe stick was 0.375 inches. To achieve the second goal–attaching the lid to the vacuum chamber–we inserted 12 through holes aligning with the threaded holes of the vacuum chamber. Lastly, to achieve the third goal–securing the sample in the center of the solenoid–we added three blind screw holes to attach the copper block to the vacuum chamber lid with brass screws.

\subsection{Copper Block}

The copper block was designed to keep the sample at a uniform temperature and to hold the sample in the center of the solenoid. For temperature control and temperature measurement, we added a cylindrical hole running through the copper block to embed a 25-ohm resistor (to serve as a heater) and to embed a Cernox sensor (thermometer). Lastly, we added three through holes so that three brass screws could connect the Vanadium sample and the PCB platform containing the Hall sensor to the copper block. 

\section{Probe Construction}

After designing the probe, we assembled it using the procedure described in this appendix. 

We first cut three notches into the disk-shaped Vanadium sample. To do so, we cut three half-circles out of the Vanadium sample, equally spaced around the circumference of the sample such that the Vanadium can fit on top of the copper block without contacting the three screws that connect the copper block to the vacuum chamber lid.

Then, we placed a Hall bar on a PCB. First, we soldered four wires to four different pins on the PCB. Next, the Hall bar was glued to the PCB using a drop of PMMA. Finally, using 1 mil aluminum wire, the four Hall bar outputs were bonded to four PCB pins that we had previously soldered wires to.

After setting up the Hall bar to the PCB, we created a mixture of equal volumes of VGE 7031 (or “g varnish”) and ethanol, and stirred until the mixture no longer produced “strings” when we dipped a stick in it. We then cut cigarette paper to the size of the top surface of the Vanadium sample, then placed the cigarette paper in the g varnish mixture and let it soak for 10 minutes. We then removed the cigarette paper from the g varnish mixture and placed it in between the Vanadium sample and the PCB, which glues the PCB to the Vanadium while ensuring that no wires from the Hall bar short to the Vanadium. 

Then, using gloves, we placed a raisin-sized dollop of N-grease on our finger and pushed the N grease into the hollow tube running through the copper block, then inserted the resistor into one end of the hollow tube in the copper block, and inserted the Cernox sensor into the other end of the hollow tube running through the copper block. 

To complete the copper block setup, we placed the Vanadium sample and the PCB on top of the copper block, then inserted three brass screws into the copper block. The screws protruded through the notches in the Vanadium sample, securing it in place.

We then connected all wires from the resistor (two wires), Cernox sensor (two wires), and Hall bar (four wires) to twisted pairs running through the inside of the probe stick.

Afterwards, we measured the resistances of all connected wires to check for shorts, correct connections, etc. We added Kapton tape to all wire connections to reduce the chance of shorting to the Vanadium, copper block, other wires, or the vacuum chamber. Once all electrical “tests” were passed, we proceeded to the next step. After each step, we re-checked all electrical connections and only moved to the next step after all electrical connection “tests” were passed.

\begin{figure}
    \centering
    \includegraphics[scale=0.1]{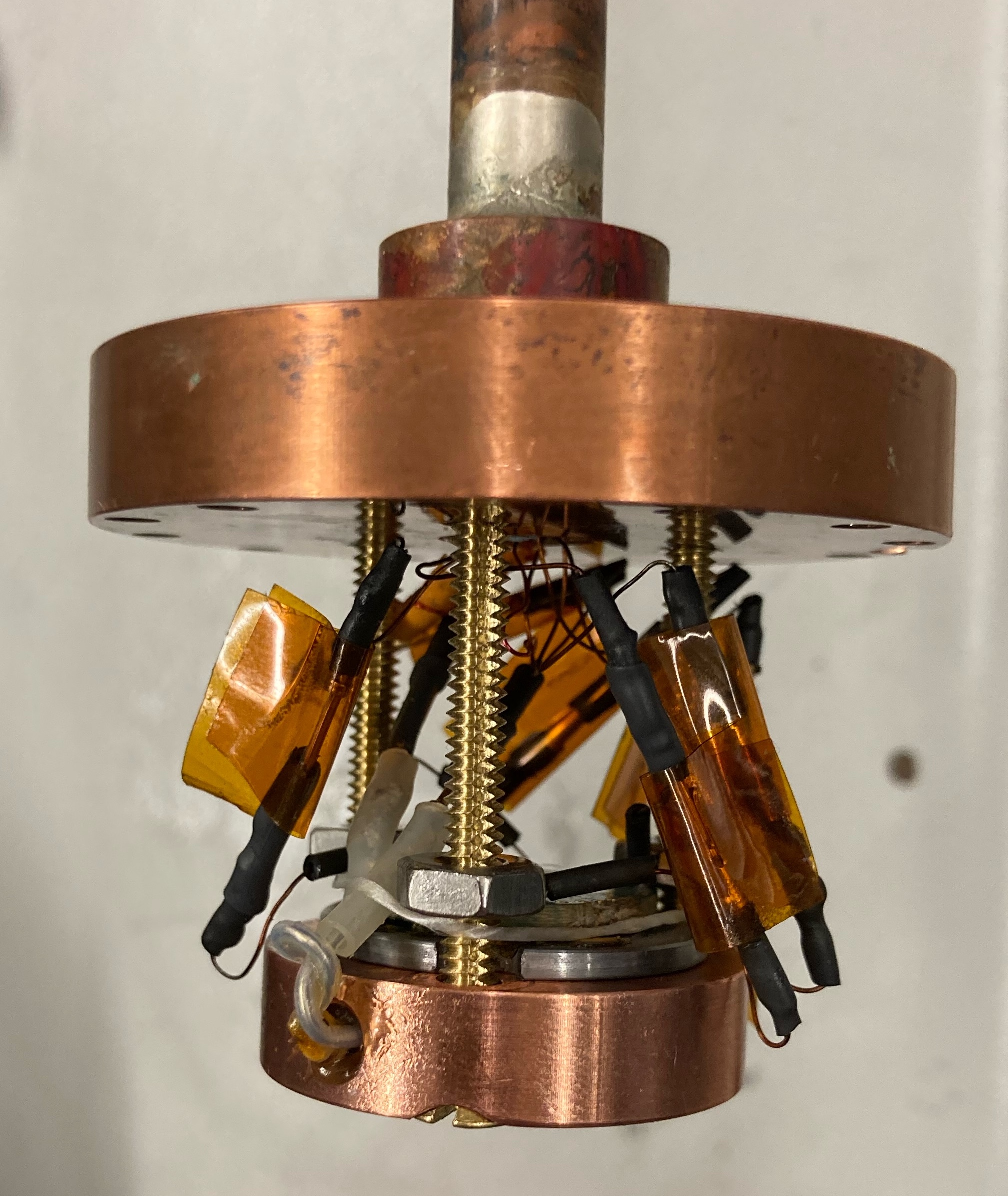}
    \caption{Photo of the probe setup before the vacuum chamber was attached to the vacuum chamber lid.}
    \label{fig:inner_probe}
\end{figure}

With the copper block setup well tested, we screwed the copper block into the lid of the vacuum chamber, then tightened the nuts on the screws, which tightened down onto the PCB, which secured the Vanadium and PCB to the copper block. This was the final step in assembling all components inside the vacuum chamber (Fig. \ref{fig:inner_probe}).

\begin{figure}
    \centering
    \includegraphics[scale=0.1]{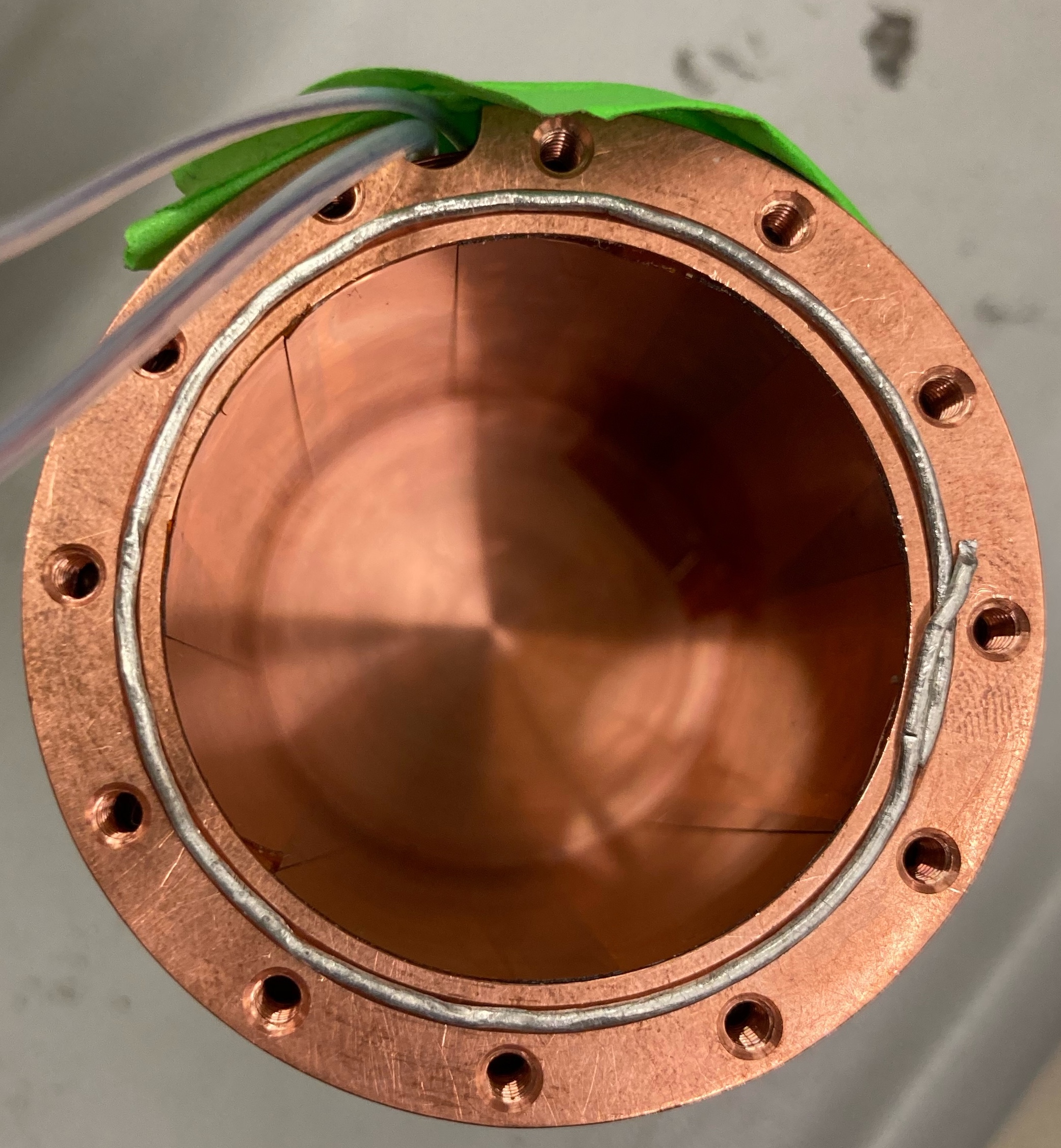}
    \caption{Photo of the vacuum chamber with indium seal aligned in the groove, ready to be secured to the lid.}
    \label{fig:indium}
\end{figure}

We applied indium seal (0.06 milli-inch diameter wire) to the groove in the vacuum chamber (Fig. \ref{fig:indium}), then placed the vacuum chamber on the lid, and secured the vacuum chamber to the lid (with the sample inside) using 4-40 brass screws in a star pattern procedure to ensure uniform compression of the indium seal.

Finally, the probe stick itself had an outer diameter of three-eights of an inch and an inner diameter of approximately one quarter of an inch. We wrapped a set of wires with a diameter of 1.2 mm in a twisted pair around the probe stick to connect the solenoid wires to the magnet leads on the other end of the probe. We further ran AWG 30 manganin and AWG 30 copper wires through the probe stick; we connected the Hall bar and Cernox sensor leads to the AWG 30 manganin wires, and we connected the AWG 30 copper wires (which are low current wires) to the resistor which we used as a heater.  

\section{Electronics}

\begin{figure}
    \centering
    \includegraphics[scale=0.7]{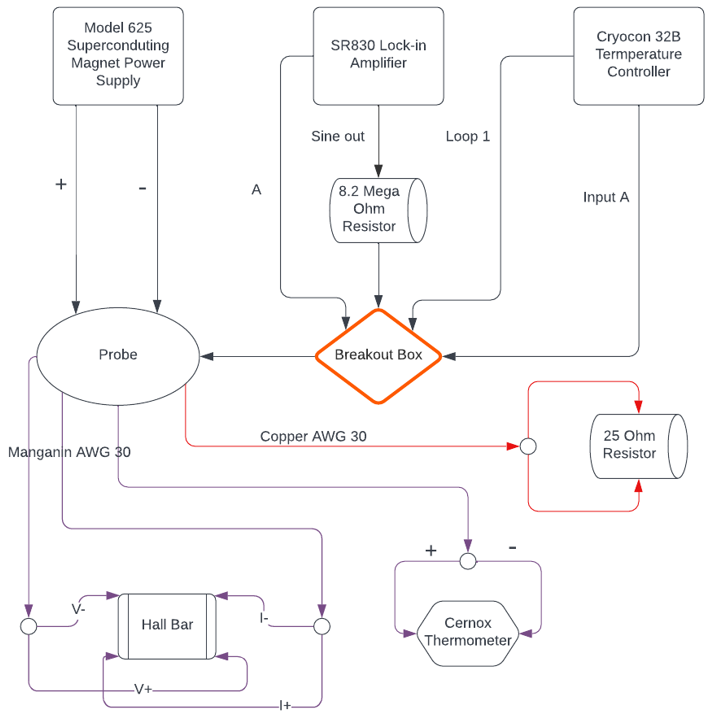}
    \caption{End to end diagram of our experiment's electronics with lines denoting wired connections. The only exception is that wires were not soldered directly onto the Hall bar, but rather onto a mounted PCB which was wire bonded to the Hall bar.}
    \label{fig:electronics}
\end{figure}

This appendix will elaborate on all electrical components shown in figure such that an attentive reader can recreate the experiment. Components will be discussed in the order they need to be connected, rather than directionally within the flowchart (Fig. \ref{fig:electronics}).

Starting with the four lines outgoing from the probe, we used a drill to twist four pairs of wire (three manganin AWG 30 pairs and one copper AWG 30 pair for the resistive heater). This is accomplished by having one team member hold both ends, while the other team member rotates a drill clamping down on the other side's ends. The pairs should have multiple twists per inch of wire once this process is completed. The four twisted pairs are then soldered to a sealed DB-9 connector that sits at the top flange of the probe. This acts as the connection between the delicate components sitting inside our vacuum chamber and our breakout box.

\subsection{Soldering}

This paragraph will briefly explain the soldering steps we employed for every connection in our experiment. Materials needed are a soldering iron, lead-free solder (preferable to avoid toxic fumes), soldering flux, a helping hand tool, needle nose pliers, heat shrink tubing, and wooden handled cotton swabs (for applying flux). Firstly, place heat shrink along the insulated portion of the wire (don't forget this step especially if one end is already soldered). The soldering iron should be heated to 700 degrees Fahrenheit if it is not a single temperature Weller. Once the iron is hot, begin applying flux with the wooden handle to the iron, wire, and solder container (for our experiment this was either a cup, as in the DB-9 connector, or a crimp contact connector, as in all connections inside our vacuum chamber). Tin the soldering iron and the wire by melting solder uniformly over the surface area. It may help to twist the wire's strands if they are loose. The tip of the iron and wire contact area should be nice and shiny before proceeding to the next step. If using a crimp contact connector, crimp the wire to the connector using the needle nose pliers before proceeding to the next step. After tinning, set up the helping hand so that the wire is resting within the solder container. Place the soldering iron close the the contact area and quickly melt solder to ensure a good electrical connection. After a couple seconds have passed and the solder has cooled, wiggle the heat shrink on top of the connection area and use a heat gun to shrink the tubing. The purpose of the heat shrink tubing is twofold: to provide insulation by covering up any exposed metal, and to provide strain relief for the wires.

\subsection{Vacuum Chamber Connections}

At this point, thread the unsoldered ends of the four twisted pairs through the probe so that the other ends can be soldered. We found that encasing the twisted pairs in Teflon tubing is quite effective in preventing wire damage during the threading. Solder all eight AWG 30 wires to a female crimp connector using the process detailed above. Solder the two ends of the Cernox thermometer to male crimp connectors, and do the same for the resistor. Attach the male and female connectors together for both ends of the Cernox and resistor, and heat shrink the connections with slightly larger tubing. The Hall Bar connections are the most delicate since they are done on tiny plated through holes (PTH's) on the PCB. Solder short wires within the same twisted pair on opposite sides of the PCB corresponding to $I_+ / I_-$ and $V_+ / V_-$. Take extra precaution in strain relieving these connections since this is likely where wires will break during the cooling process (as shown in Figure E.2), if any. Solder male connectors on the end of these short wires, and attach to the remaining female connectors once completed. Apply generous heat shrink to these connections as with the Cernox and resistor connections. Once these small connections are done, the vacuum chamber is ready to be fully assembled.

\begin{figure}
    \centering
    \includegraphics[scale=0.06, angle=270]{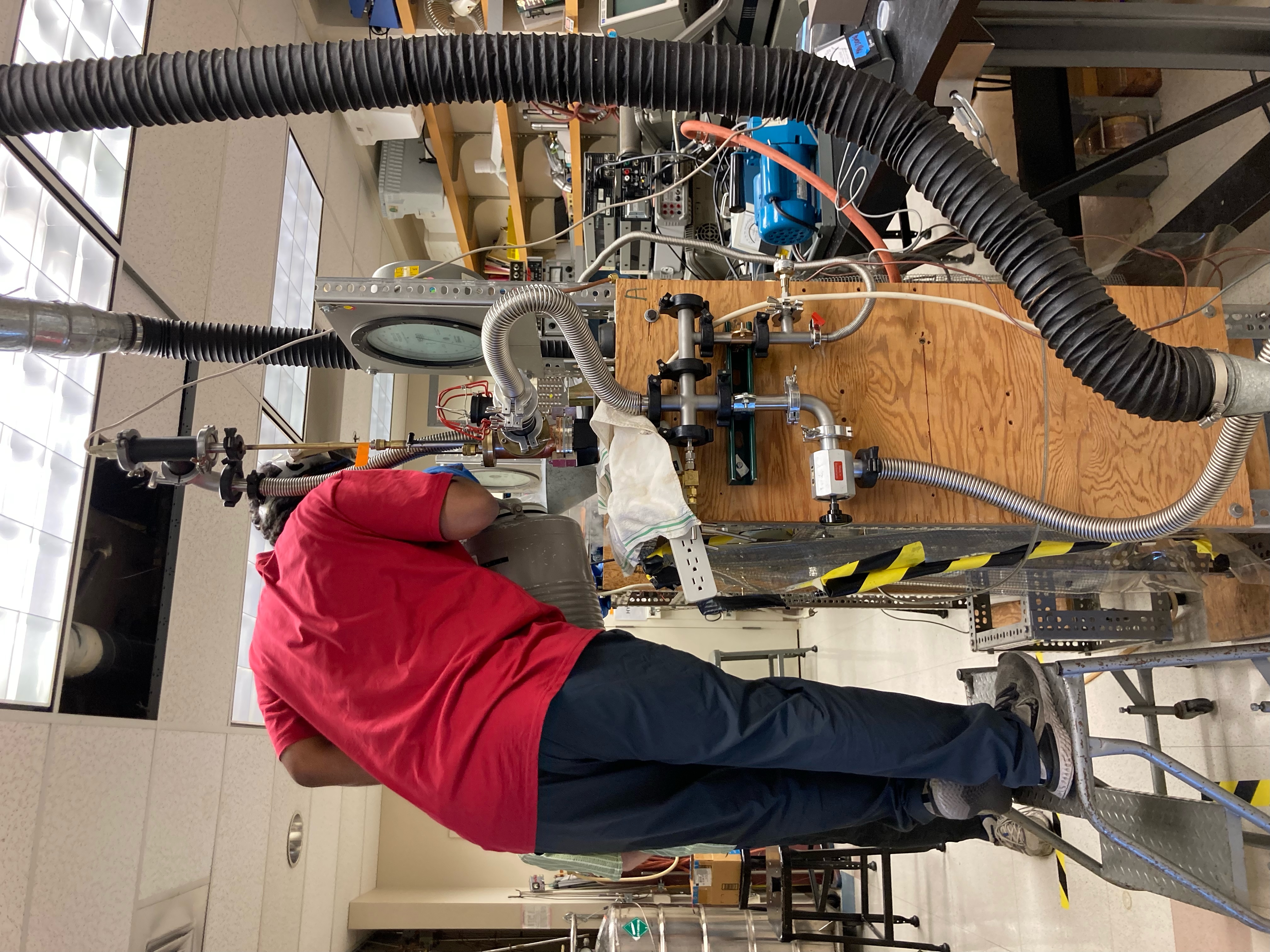}
    \caption{Cooling our sample to liquid nitrogen temperature.}
    \label{fig:electronics}
\end{figure}

\subsection{External Connections}

To wire up the solenoid to the current leads (as shown in Figure E.3), begin by soldering the ends of the solenoid wire to female crimp contact connectors. Make sure the connectors are large enough to fit 1.2mm diameter wires although the solenoid wires are much thinner (since this is the wire size that connects to the current leads). Solder the twisted pair 1.2mm wires to male crimp contact connectors. Attach these two connectors together once the chamber is fully assembled to prepare for data collection.

\begin{figure}
    \centering
    \includegraphics[scale=0.06, angle=180]{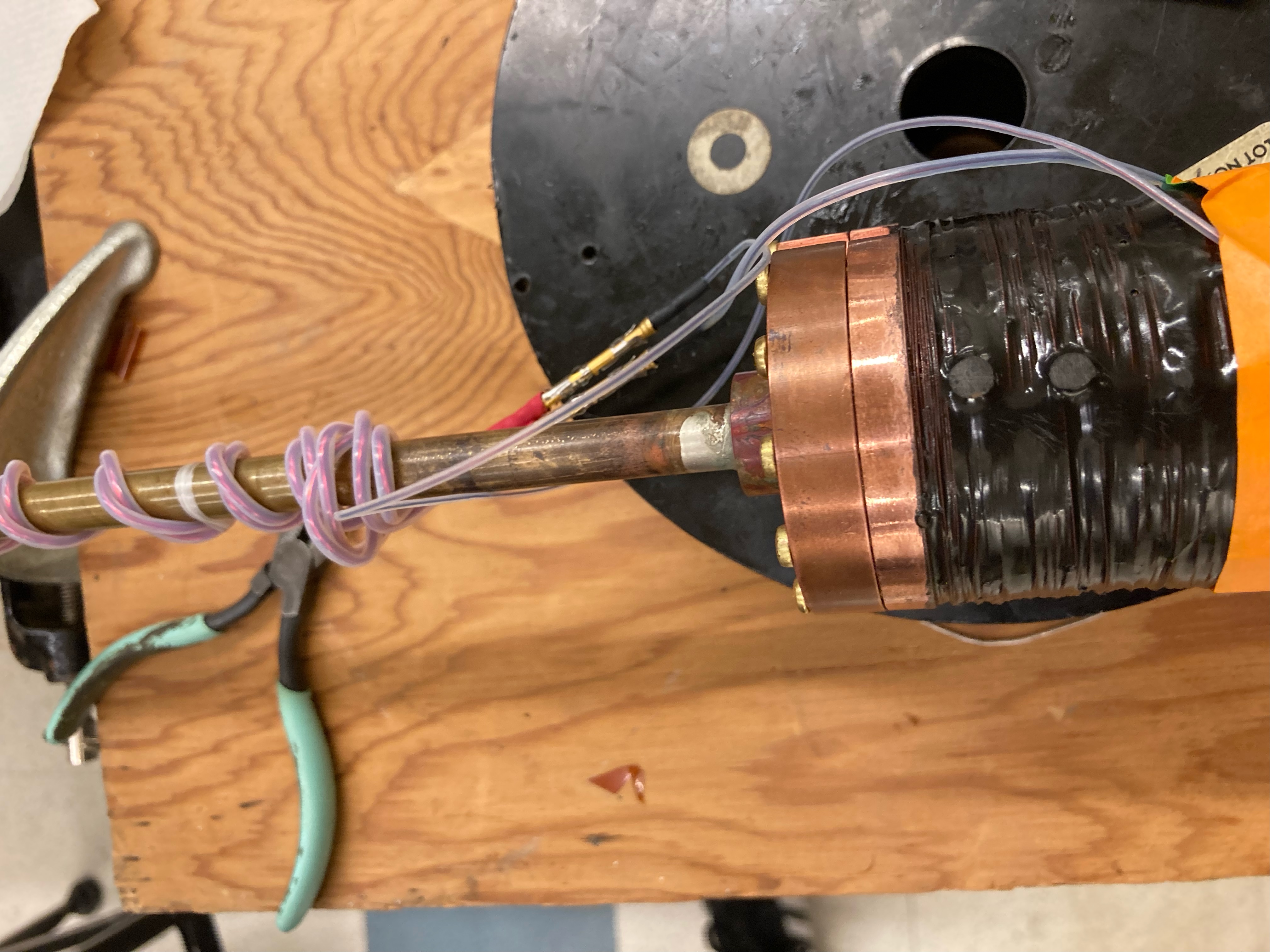}
    \caption{Outside view of solenoid and connecting magnet leads.}
    \label{fig:electronics}
\end{figure}

Transitioning to wiring outside of the probe, we took a multicore cable that connected to our DB-9 and fed that as input to our breakout box (shown in figure E.4). The purpose of the breakout box is to keep the wiring shielded and organized while providing an extra layer of strain relief (same dual purpose as the heat shrink). At this stage make sure you label which wires each component corresponds to, which we did by making a pin layout of both the inside and outside DB-9's using the different colors of our multicore cable wires (remember the connections get mirrored on opposite sides or the flange). We soldered the four Hall bar wires to two female BNC socket connectors (for a single connector, one wire goes to the inside, the other to the outer shield). We did the same BNC socket connection for the two wires coming from the resistive heater. Soldering the two wires stemming from the Cernox thermometer was more challenging because this is typically a 4-terminal measurement. To shrink the measurement to 2-terminal, we connect the $V_+/V_-$ wires in parallel to the $I_+/I_-$ wires. Specifically, we soldered the two Cernox ends to another multicore cable that fed as input to a 6 pin DIN connector (this input is required for the Cryocon and made the electronics component much harder due to space constraints). The two wires from the multicore cable were soldered to pins 1 and 5 of the pinout, and jumpers were soldered from these pins to pins 2 and 4, respectively (the $V_+/V_-$ pins). As with the PCB connections, keep in mind that all wires must fit within the tiny DIN connector when cutting wires/heat shrink tubing.

\begin{figure}
    \centering
    \includegraphics[scale=0.06, angle=180]{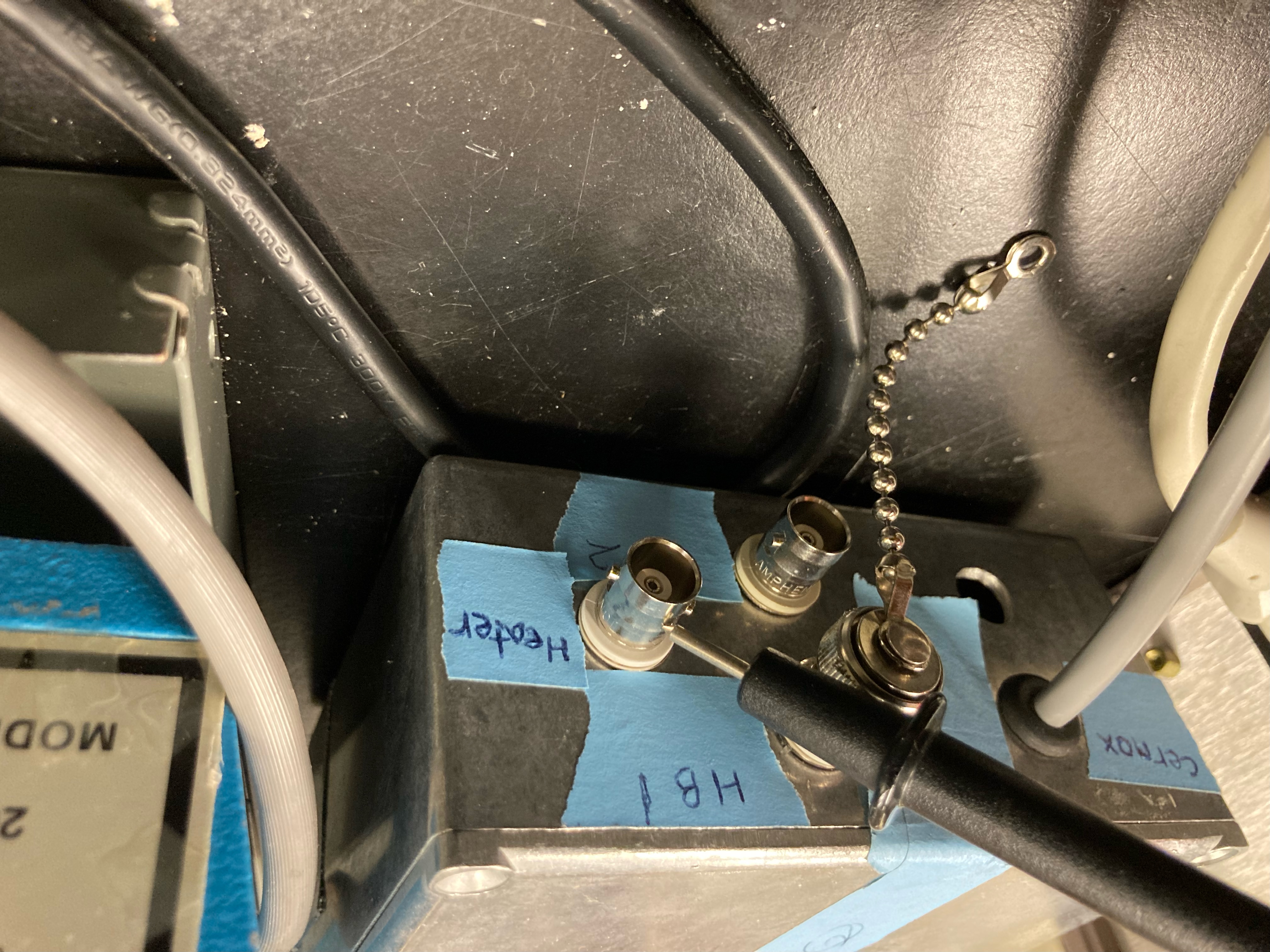}
    \caption{Outside view of breakout box during electrical connection testing during cooldown.}
    \label{fig:electronics}
\end{figure}

The last miscellaneous connection is soldering the $8.2 M \Omega$ resistor that transforms the Lock-in from a voltage source to a current source. We soldered both ends of this resistor to BNC socket connectors and encased the resistor in its own box topped off with copper tape.

Once all the soldering was finished, we connected all of the wires to our 625, SR830, and Cryocon 32B. The 625 is electronically simple, though it requires wires with little resistance connecting to the female banana sockets of our superconducting magnet (little resistance is better to prevent the power supply from detecting a quench; we had to swap out high resistance wires because our power supply was detecting quenches during data collection at sub liquid nitrogen temperatures). The SR830 needs three BNC's: a short one connecting "sine out" from the amplifier to one end of the $8.2 M \Omega$ resistor, another connecting the other side of the $8.2 M \Omega$ resistor to one BNC socket on the breakout box, and the last one connecting the other BNC socket to "A" on the Lock-in. The 6 pin DIN is connected to the back of the Cryocon 32B (feeds into "input A") along with a double banana plug attached to the last BNC from the breakout box (feeds into "Loop 1"). Lastly, the 625, SR830, and Cryocon 32B should all be daisy chained by GPIB cables for interfacing during data collection.

\section{Software-Hardware Interface}

This appendix will detail the scripts used during data collection to interface with our electronics. The additional hardware required is a National Instruments GPIB to USB driver. To use the driver, we first downloaded the National Instruments software. Using the software, we configured each device's communication parameters according to their manuals (mode, parity, baud rate, etc.). Once each device was configured, we used the software to identify each devices GPIB address, a piece of information necessary for interfacing with the device using Virtual Instrument Software Architecture (VISA device). 

Our experiment was interfaced by four scripts, all located in the folder "Desktop/108 Team 3" in the computer on our experiment table. Each device was operated by its own script, and the final script served as the higher level facilitator script of the experiment. This higher level script did nothing more than initializing the devices objects and running their corresponding scripts.

The Cryocon 32B script was the simplest of the lot and ultimately wasn't used during data collection. It was intended to remotely control the temperature of our sample but had some bugs that we deemed not worthy of sorting through. We decided to manually set the temperature on the Cryocon because temperature was our slow axis, requiring only a few setpoints for our entire experiment. This decision also allowed us to sanity check each component before ramping our applied field instead of hoping our electrical components would work throughout the entire course of the experiment (see note about quenching in above index; it did not work throughout the entire course of our experiment).

The Model 625 script was responsible for initializing power supply parameters such as our coil constant, max ramp rate, and telling the machine we have no persistent switch (which is not necessary because our power supply has a built in stabilizing resistor). It also conducted ramping our external magnetic field over a symmetric interval (i.e. $[-0.1T, 0.1T]$, range depends on the sample temperature since lower temperatures require a larger range to distinguish phase diagram regions), initially ramping to one of the extremes like $0.1 T$. This script also took in a parameter $N$ indicating the number of measurements we wanted to make throughout the ramping process. We scaled this parameter $N$ so that we would be taking data over $2 mT$ intervals. We found this resolution sufficient for determining the region boundaries, though increasing this parameter could help us make a more accurate phase diagram in future work. The quantities we recorded during each measurement were the external current, external magnetic field, Hall voltage, and sample temperature. The Hall bar magnetic field is computed after the entire ramping process, since an anti-symmetrization is required for obtaining an accurate plot (details provided in Hall bar appendix). The script also ramped the external magnetic field down to $0 T$ after all data was collected so that the device could be powered off properly.

The Lock-in script starts by setting the output voltage to $0.4 V$ (which corresponds to around $50 nA$ using the $25 M \Omega$ resistor). We found this value to yield a cleaner Hall voltage reading in practice (as opposed to our initial $0.2 V$). The script then implements autoscaling of the Hall voltage reading so that data collection is as automated as possible. It also reads all of our data described previously and stores it to a file after every interval. It finally takes a parameter $d$, indicating the number of seconds to delay before live plotting the Hall voltage vs external applied field.

\section{Magnetic Field Simulations}

Since the Hall sensor is not directly on the surface of the Vanadium disk sample, even in the Meissner phase the Hall sensor will detect a non-zero magnetic field. Outside the Vanadium disk, inside the vacuum chamber it is a reasonable assumption that there is no appreciable concentration of net positive or net negative charge. There are small currents being sent through wires, which will produce small magnetic fields. However, the fields being sensed by the Hall sensor will be of the order of about one-tenth of the external magnetic field in the presence of no currents, as will be shown shortly. This is much to small as to be affected by the minute currents, so we can discount the existence of these currents. Note that this is a reasonable assumption because we are a bit away from the surface of the Vanadium sample; if the Hall sensor measured the magnetic field right at the Vanadium disk itself, then the Hall sensor would predominantly register the magnetic field being produced by the small currents themselves (plus noise).

The solenoid produces an external magnetic field which has a fixed direction and magnitude. I will suppose that this direction is pointed vertically upwards, in the direction of the positive z-axis. The disk is perpendicular to the z-axis, centered at the origin. 

Therefore, we can assume that in the vicinity of the Vanadium disk, outside the disk and inside the vacuum chamber, there are no currents and charges. The curl and divergence of the magnetic field is zero, which means that the magnetic field equals negative the gradient of a scalar potential which satisfies Laplace's equations. We thus can apply Laplace's equation with suitable boundary conditions to calculate the magnetic field around the disk. Since the Vanadium disk expels all magnetic flux, the magnetic field perpendicular to the surface of the disk must be zero. This specifies a Neumann boundary condition. Further, we know that far away from the disk, the magnetic field should equal approximately the external magnetic field. This specifies a second Neumann boundary condition. Finally, note that adding any constant to the scalar potential leaves the magnetic field invariant. We can eliminate this ambiguity by fixing the value of the potential at an equipotential surface. Far away from the disk, the magnetic field is of constant magnitude and pointed in the z-direction, so the equipotential surfaces are planes of constant z-value. I can set the potential at some arbitrary plane to equal zero. 
With these conditions, we can use Mathematica's finite element solver to compute the scalar potential and hence the magnetic field in the vicinity of the magnetic. Some remarks on numerical simulation are warranted here. First, because Mathematica's free cloud finite element partial differential equation solver is a bit brittle, we decided to make the following reductions to ensure the simulation was accurate. First, we first exploited axial symmetry around the z-axis to reduce Laplace's equation to a two dimensional problem involving only the z coordinate and the cylindrical radius coordinate. Second, to impose that the magnetic field far away from the disk is pointed in the z direction and is of a constant fixed magnitude, we set the magnetic field at an order of thirty centimeters from the disk to equal the external magnetic field. Specifically, we specified a cylindrical region as an external boundary in addition to the disk's surface as an internal boundary to specify the region in which we would be solving Laplace's equation. We found that having the cylindrical boundary be too far results in the numerical solver having a discretization of spacetime which is not finely grained enough to capture the magnetic field at the scales of the disk, while having the cylindrical boundary be too small (i.e., the boundary being too close tot he cylinder) results in numerical distortions (because the magnetic field is set to a fixed value just out of the cylinder, resulting in the magnetic field distribution being strongly distorted to quickly reach this external value). We ended up settling for the cylindrical radial coordinate to be in a range from $0$ to $200$ and for the z-coordinate to range from $-100$ to $100$, as this produced decent results. 

With these specifications in hand, we ran Mathematica's finite element solver to compute the magnetic scalar potential. Here is a plot of the z-component of the magnetic field in the vicinity of the Vanadium disk, as a function of the height above the top of the Vanadium disk (with the cylindrical radius coordinate set equal to zero). Not that I here set the external magnetic field to equal $1T$.

\begin{figure}
    \centering
    \includegraphics[scale=0.3]{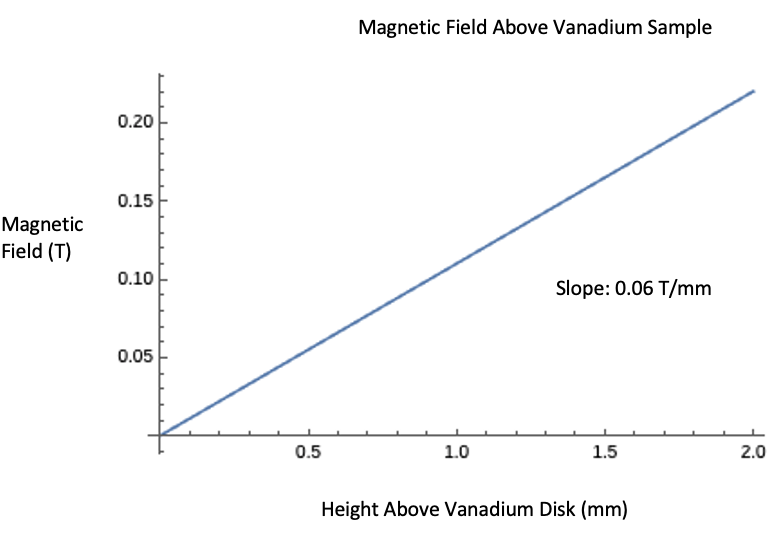} 
    \caption{Plot of magnetic field as a function of height above the center of the Vanadium disk.}
    \label{fig:scatter_plot_2.6}
\end{figure}

As this plot shows, near the Vanadium disk the magnetic field varies linearly as a function of height. At a height of $1.5 mm$, which is approximately where we placed our hall bar, the magnetic field is about $0.09T$. Hence, if we change the external magnetic field from $1T$ to $B_{ext}$, then by the linearity of Laplace's equation the magnetic field measured by the hall bar at a distance of $1.5 mm$ from the disk is $0.09 B_{ext}$. Hence, the Meissner phase will produce a shallow line for a plot of the hall bar magnetic field versus the external magnetic field, with a slope of around $0.09$.

\section{Hall Bar}

\subsection{Calculating the Hall Coefficient}

On a particular charge carrier in the Hall sensor, the electric and magnetic forces must balance in equilibrium. Hence, $vB = E$, where I assume the magnetic field is perpendicular to the Hall sensor's surface and the electric field is tranverse to the direction of current flow. Now, $E = \frac{V_H}{w}$, and $v = \frac{I}{nwe}$. Hence, we get that $\boxed{V_H = \frac{I}{ne} B}$. 

In particular, with $I = 10^{-7} A$ and $n = 5 \cdot 10^{11} cm^{-2}$, we get that $\frac{V_H}{B} = 12.5 \frac{\mu V}{T}$. Hence, a resolution of $10^{-3} T$ for the magnetic field requires a resolution of $12.5 nV$ for measuring the Hall voltage.

\subsection{Anti-symmetrizing the Hall Bar Measurement}

The current flow through the Hall bar will not be exactly perpendicular to the direction of the Hall voltage, i.e. $V_{xy}$ where $x$ and $y$ denote the two surfaces of the Hall sensor. This occurs for a variety of reasons, from irregularities in the current to the equipotential surfaces not being exactly perpendicular to the $x$ and $y$ surfaces. Because of this, when we measure the Hall voltage, we will in fact be getting a mixture of $V_{xy}$ (our desired voltage) with a small bit of $V_{xx}$. 

Now, the $V_{xy}$ voltage is odd while the $V_{xx}$ voltage is even. That the $V_{xy}$ voltage is odd follows from the hall effect (fron which we can derive that $V_{xy}$ is proportional to the external magnetic field). That $V_{xx}$ is even arises from the fact that $B$ affects $V_{xx}$ primarily by affecting the resistance the current facing while going through the hall bar surfaces; how $B$ should change the resistance of the hall bar should not depend on the sign of $B$.

Hence, we can interpret the statement that the measured hall voltage $V_H$ is a mixture of the desired $V_{xy}$ voltage with a bit of the $V_{xx}$ voltage in the following way:our $V_H$ reading will be a function with both odd and even components, with the odd components dominating. We want to get rid of the even components corresponding to $V_{xx}$, which we can do with the following transformation: $V_H(B) \rightarrow \frac{V_H(B) - V_B(-B)}{2}$. This transformation isolates the odd component, as the following argument illustrates: if a function $f(x)$ equals the sum of an even and an odd component $f_e(x) + f_o(x)$, then $\frac{f(x) - f(-x)}{2} = \frac{f_e(x) + f_o(x) - f_e(-x) - f_o(-x)}{2} = \frac{f_e(x) + f_o(x) - f_e(x) + f_o(x)}{2} = \frac{2f_o(x)}{2} = f_o(x)$. Hence, when processing the Hall bar measurement, once we have a table of the Hall bar measurements as a function of magnetic field, we should anti-symmetrize the hall bar measurements to isolate the desired $V_{xy}$ voltage. 
 
\section{Thermometer Calibration}

To calibrate and interpret the thermometer data over a continuous region, we explored two options, each starting with obtaining the thermometer's resistance measured at different temperatures. 

For the first approach, we graph the data points on a resistance versus the temperature plot. Afterwards, we split the temperatures into three regions: 2-10 K, 10-50 K, 50-300 K. In each region, we fit $\frac{1}{T}$ to a cubic polynomial in terms of $\mbox{ln}(R)$ (this is basically an application of the Steinhart-Hart equation with the second order term intact). We should make sure that the regression plots at consecutive temperature ranges meet at the same point (so that the overall regression line for resistance given temperature is continuous).  

The second approach applies a quadratic interpolation over the $T(R)$ data, which yields a slightly less accurate fit compared to the triple region approach over our region of interest (2K - 10K is our region of interest since this is the range where all of our measurements were taken). Increasing the non-linearity of the interpolation had diminishing returns, which is why we opted for the first approach in analyzing our calibrated thermometer.

This calibration curve can then be inserted into the Cryocon by using their interfacing software. We tried interfacing with their software using the National Instruments driver, but this GPIB connection failed to no end. As a backup plan, we used an RS-232 connector to upload the calibration curve to the device. Make sure the resistance is specified as $\log \Omega $'s in the software and that the curve does not include too many points, otherwise the device will not accept it.

\section{Temperature Control}

Heat can be transferred by three different mechanisms: conduction (through molecular contact), convection (through a fluid), and radiation (by E-M waves) which we take as negligible relative to the other mechanisms of heat flow \cite{Ekin2006} . Therefore, the two most significant sources of temperature decrease of the vacuum chamber was due the stainless steel screws connecting the copper vacuum chamber lid to the copper block that holds the sample (conduction) and the the presence of the exchange gas, Helium, in the vacuum chamber (convection). The main source increasing the temperature is the power dissipated by the resistor embedded in the copper block (conduction).

\subsection{\label{app:subsec}Quantifying the heat flow through exchange gas}
Here, we calculate the heat flow through the exchange gas in the vacuum chamber. The heat flow between two surfaces with temperature difference $\Delta T$ with a low-pressure exchange gas (free-molecule case) between the two surfaces, is given in Watts by, 
$$ \dot{q}_{gas} = ka_0PA_i \Delta T$$
where $P$ is pressure in Pascals and $\Delta T$ is the temperature difference in Kelvin \cite{Ekin2006}. Notice that $\dot{q}_{gas}$ is independent of the separation distance between the two plates since the mean free path is on the same scale as the separation distance of the plates because the pressure of the exchange gas is low. When the geometry of the system is two parallel plates, $a_0 \approx 1/3$; for Helium, $k = 2.1$, and $A_i$ is the surface area, in units $m^2$, of both parallel plates, where here both surfaces have the same surface area $A_i = A = \pi r^2 \approx 10^{-3} m^2$. Therefore, the heat flow due through the exchange gas, in Watts, is,

$$ \dot{q}_{gas} = 10^{-3} P \Delta T$$

so to get an upper-bound estimate on the cooling due to the exchange gas, we take a pressure of $100$ Pascals and a temperature difference of $3K$, giving $\dot{q}_{gas} \approx 10^{-1}$ Watts. 

\subsection{\label{app:subsec}Quantifying the heat flow through the stainless steel connection between the vacuum chamber and the copper block}
From Ekin: "the conduction heat flow $\dot{q}_{cond}$ through" a solid bar of length $L$ is given by 
$$ \dot{q}_{cond} = k(T) A \Delta T / L$$
where $A$ is its cross-sectional area, $\Delta T$ is the temperature difference across the bar, and $k(T)$ is the temperature-dependent thermal conductivity of the material. For stainless steel, we approximate $k(T) \approx k(T=4K) \approx 2 \cdot 10^{-1} W/(m \cdot K)$. We estimate cross-sectional area of the stainless steel $A \approx 10^{-5} m^2$, and estimate the length of the screw $L \approx 10^{-2} m$. Therefore, the heat flow is, $$ \dot{q}_{cond} = 2 \cdot 10^{-4} \Delta T.$$ So, for an upper-bound estimate, we take $\Delta T = 3K$  which gives an upper-bound heat flow estimate, 
$$ \dot{q}_{cond} = 1.8 \cdot 10^{-3} \text{Watts}$$
where we multiplied by a factor of three since there are three stainless steel screws connecting the vacuum chamber to the copper block. 

\subsection{\label{app:subsec}Quantifying the radiative heat transfer}
The heat transfer due to radiation between two surfaces is given in Ekin by, 
$$ \dot{q}_{rad} = \sigma EA({T_2}^4 - {T_1}^4)$$ where $\sigma$, the Stefan-Boltzmann constant, is $5.67 \cdot 10^{-8} W/(m^2K^4)$, A is a factor that depends on the geometry, and subscripts 1 and 2 of $T$ refer to cold and warm surfaces, respectively. We approximate the system, a copper platform surrounded by a copper vacuum chamber, with the concentric-sphere formula (a valid approximation, according to Ekin). In this case, the area $A$ is that of the enclosed surface (where we estimate $A \approx 2 \cdot 10^{-3}m^2$, and our material is best categorized by the diffuse reflection formula, so $E = \varepsilon_1 \varepsilon_2 / (\varepsilon_1 + \varepsilon_2 - \varepsilon_1 \varepsilon_2) \approx 0.05$, since $\varepsilon_1 \approx \varepsilon_2 \approx 0.1$. To estimate the upper-bound heat transfer due to radiation, we take $T_1^4 = 4K,T_2^4 = 7K$, and plug in to get an upper-bound radiative heat transfer estimate,  $\dot{q}_{rad} = 1.8 \cdot 10^{-10}$ Watts. This heat transfer is orders of magnitude less than the upper-bound heat transfer estimates due to conduction (due to connection through stainless steel rods) and convection (due to Helium exchange gas in vacuum chamber). 

\subsection{\label{app:subsec}Total cooling power estimation}
We have shown above that our upper-bound estimates of heat transfer due to exchange gas, stainless steel conduction, and radiation to be $10^{-1}$ Watts, $1.8 \cdot 10^{-3}$ Watts, and $1.8 \cdot 10^{-10}$, respectively. Therefore, the upper-bound of the total cooling power is their sum, which is approximately $10^{-1}$ Watts.

\subsection{\label{app:subsec}Quantification of power dissipated by the resistor as a function of current}
We used a $25 \Omega$ resistor that can dissipate a maximum of $0.25$ Watts. Since the power dissipated by the resistor is given by $P = I^2 R$, where $I$ is the current through the resistor and $R \approx 25 \Omega$, then we can tune the heat transfer from the resistor to the copper platform by tuning the current running through the resistor. Since the maximum power that the resistor can dissipate is $0.25$ Watts, which is greater than the upper-bound estimate of the total cooling power, we conclude that we have the ability to both heat the sample or cool the sample.

\section{Additional Information in Data Analysis}

We plotted the regression lines from the normal phase, imposed on a scatter plot of data points of the hall bar's magnetic field versus external magnetic fields for each temperature we measured (Figures K1-8). 

\begin{figure}[H]
    \centering
    \includegraphics[scale=0.5]{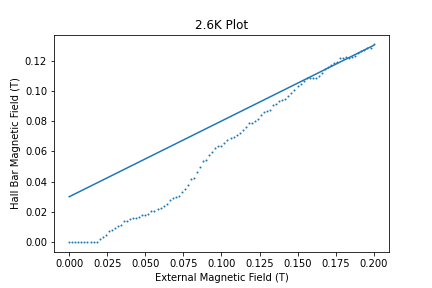}
    \caption{Scatter plot at 2.6K of the magnetic field of the hall sensor versus the external magnetic field.}
    \label{fig:scatter_plot_2.6}
\end{figure}

\begin{figure}[H]
    \centering
    \includegraphics[scale=0.5]{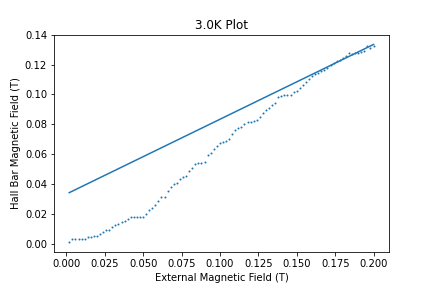}
    \caption{Scatter plot at 3.0K of the magnetic field of the hall sensor versus the external magnetic field.}
    \label{fig:scatter_plot_2.6}
\end{figure}

\begin{figure}[H]
    \centering
    \includegraphics[scale=0.5]{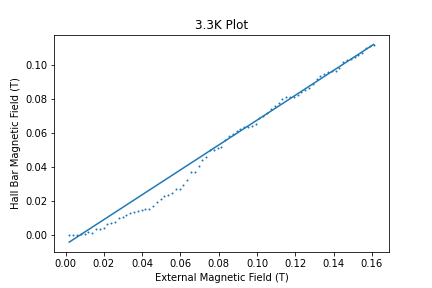}
    \caption{Scatter plot at 3.3K of the magnetic field of the hall sensor versus the external magnetic field.}
    \label{fig:scatter_plot_2.6}
\end{figure}

\begin{figure}[H]
    \centering
    \includegraphics[scale=0.5]{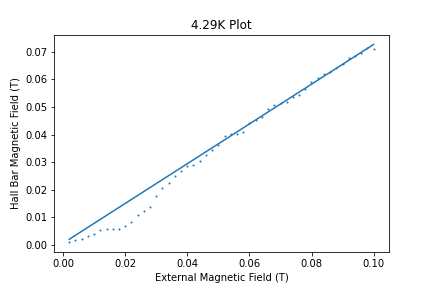}
    \caption{Scatter plot at 4.29K of the magnetic field of the hall sensor versus the external magnetic field.}
    \label{fig:scatter_plot_2.6}
\end{figure}

\begin{figure}[H]
    \centering
    \includegraphics[scale=0.5]{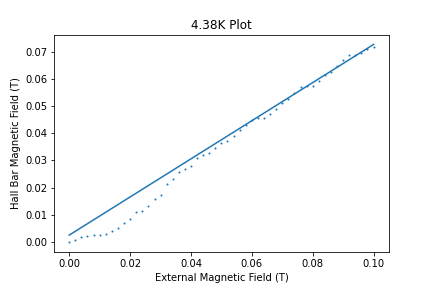}
    \caption{Scatter plot at 4.38K of the magnetic field of the hall sensor versus the external magnetic field.}
    \label{fig:scatter_plot_2.6}
\end{figure}

\begin{figure}[H]
    \centering
    \includegraphics[scale=0.5]{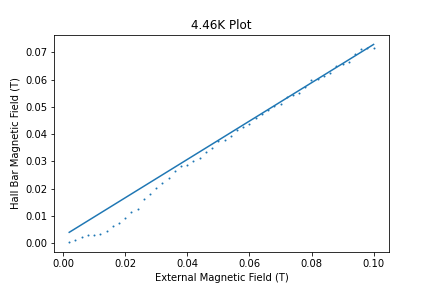}
    \caption{Scatter plot at 4.46K of the magnetic field of the hall sensor versus the external magnetic field.}
    \label{fig:scatter_plot_2.6}
\end{figure}

\begin{figure}[H]
    \centering
    \includegraphics[scale=0.5]{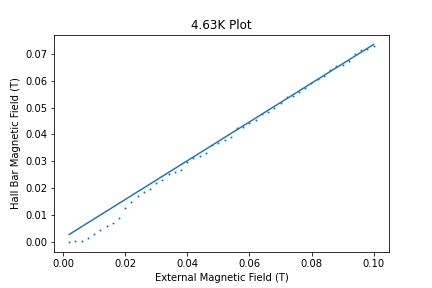}
    \caption{Scatter plot at 4.63K of the magnetic field of the hall sensor versus the external magnetic field.}
    \label{fig:scatter_plot_2.6}
\end{figure}

\begin{figure}[H]
    \centering
    \includegraphics[scale=0.5]{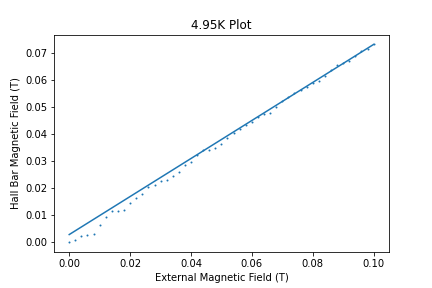}
    \caption{Scatter plot at 4.95K of the magnetic field of the hall sensor versus the external magnetic field.}
    \label{fig:scatter_plot_2.6}
\end{figure}


\end{document}